\documentclass[twocolumn, journal]{IEEEtran}
\IEEEoverridecommandlockouts

\usepackage{diagbox}
\usepackage{bm}
\usepackage{color}
\usepackage{graphicx}
\usepackage{amsmath}
\usepackage{amssymb}
\usepackage{algorithm}
\usepackage{algorithmic}
\usepackage{ulem}
\usepackage{lineno}
\usepackage{lettrine}
\usepackage{subfigure}
\usepackage{epstopdf}
\usepackage{stfloats}
\usepackage{color}
\usepackage{graphicx}
\usepackage{amsmath}
\usepackage{amssymb}
\usepackage{algorithm}
\usepackage{algorithmic}
\usepackage{amsmath}
\usepackage{multirow}
\usepackage{booktabs}
\usepackage{array}
\usepackage{amsthm}
\usepackage{stfloats}
\usepackage{caption}
\usepackage{bm}
\usepackage{booktabs}
\usepackage{setspace}
\usepackage{makecell}
\usepackage{cases}

\newcommand{\eqdef}{\triangleq}
\newcommand{\be}{\begin{equation}}
\newcommand{\ee}{\end{equation}}
\newcommand{\bea}{\begin{eqnarray}}
\newcommand{\eea}{\end{eqnarray}}
\newcommand{\ba}{\begin{array}}
\newcommand{\ea}{\end{array}}
\newcommand{\nid}{\noindent}

\newcommand{\bolds}{\boldsymbol}

\makeatletter

\renewcommand{\maketag@@@}[1]{\hbox{\m@th\normalsize\normalfont#1}}%

\makeatother

\title{Cram\'{e}r-Rao Bound Optimization for Active RIS-Empowered ISAC Systems
\thanks{Q. Zhu and M. Li are with the School of Information and Communication Engineering, Dalian University of Technology, Dalian 116024, China (e-mail: qzhu@mail.dlut.edu.cn; mli@dlut.edu.cn).}
\thanks{R. Liu is with the Center for Pervasive Communications and Computing, University of California, Irvine, CA 92697, USA (e-mail: rangl2@uci.edu).}
\thanks{
Q. Liu is with the School of Computer Science and Technology, Dalian University of Technology, Dalian 116024, China (e-mail: qianliu@dlut.edu.cn).}
}

\author{ \IEEEauthorblockN{Qi Zhu, Ming Li, \textit{Senior Member, IEEE}, Rang Liu, \textit{Member, IEEE},\\ and Qian Liu, \textit{Member, IEEE}}}

\begin{document}
\maketitle
\thispagestyle{empty}

\begin{abstract}
  Integrated sensing and communication (ISAC), which simultaneously performs sensing and communication functions within a shared frequency band and hardware platform, has emerged as a promising technology for future wireless systems. Nevertheless, the weak echo signal received by the low-sensitivity ISAC receiver significantly constrains sensing performance in scenarios involving obstructed targets. Active reconfigurable intelligent surface (RIS) has become a prospective solution by situationally manipulating the wireless propagations and amplifying the signals. In this paper, we investigate active RIS-empowered ISAC systems to enhance radar echo signal quality as well as communication performance. In particular, we focus on the joint design of the base station (BS) transmit precoding and the active RIS reflection beamforming to optimize the parameter estimation performance in terms of Cram\'{e}r-Rao bound (CRB) subject to the communication users' signal-to-interference-plus-noise ratio (SINR) requirements. An efficient algorithm based on alternating optimization, semidefinite relaxation (SDR), and majorization-minimization (MM) is proposed to solve the formulated challenging non-convex problem. Finally, simulation results validate the effectiveness of the developed algorithm and the potential of employing active RIS in ISAC systems to enhance direct-of-arrival (DoA) estimation performance.
\end{abstract}


\section{Introduction}

Both high-quality wireless connectivity and high-accurate sensing ability are required in next-generation wireless networks to support intelligent manufacturing, intelligent transportation, intelligent medical, and other emerging applications.
Benefiting from the widespread employment of millimeter wave (mmWave) and massive multiple-input multiple-output (MIMO) technologies, communication signals exhibit high resolution in the angular domain, enabling sensing with communication signals, which motivates integrated sensing and communication (ISAC) to become one of the leading technical trends \cite{LiuF}-\cite{LiuA}. By utilizing a fully shared platform and transmitting dual-functional waveforms to simultaneously perform sensing and communication, ISAC is expected to significantly boost spectral efficiency and energy efficiency as well as reduce hardware costs and signaling overhead.

Instead of treating communication and sensing as two separate goals, ISAC pursues a deeper level of integration of them to achieve mutual benefit through co-designs. Various advanced signal processing techniques are proposed to design dual-functional waveforms based on different communication/sensing metrics \cite{LiuX}-\cite{LiuR1}.
Typical communication performance metrics include the signal-to-interference-plus-noise ratio (SINR), the multi-user interference (MUI), and the achievable sum-rate. Meanwhile, frequently used sensing performance metrics include the beampattern mean squared error (MSE), the waveform similarity, and the received echo power/signal-to-noise ratio (SNR). In addition to the metrics mentioned above, Cram\'{e}r-Rao bound (CRB) is also an essential sensing performance metric for target parameter estimation, which provides a lower bound on the variance of any unbiased estimator \cite{LiuF4}, \cite{HuaH}.

Reconfigurable intelligent surface (RIS) is also recognized as a key enabling technology for next-generation wireless networks thanks to its superior ability to intelligently reconfigure the wireless propagation environment. Specifically, RIS is a meta-surface composed of passive electromagnetic elements, in which each element can individually and adaptively tune the phase-shifts of the incident signals. By intelligently coordinating reflections, RIS can establish a favorable virtual line-of-sight (LoS) link between the transmitter and the receiver, thus providing a novel approach for addressing wireless channel fading impairment and interference problems \cite{Basar}, \cite{Wu}. By introducing additional degrees of freedom (DoFs), deploying RIS in existing wireless networks can not only substantially improve communication quality, but also significantly expand communication coverage \cite{ZhouG}-\cite{ZengS}.

The success of RIS applications in various communication scenarios has inspired research to explore the combination of RIS and ISAC technologies \cite{LiuR2}-\cite{Luo2}. The authors in \cite{JiangZ}, \cite{YanS} employ RIS within ISAC systems to offer additional propagation paths for radar echo signals and simultaneously enhance communication performance. In \cite{SongX}, RIS is utilized to create virtual LoS links to sense the potential target blocked by obstacles. The co-design of transmit waveform, receive filter, and RIS reflection coefficients in the presence of strong clutter in the RIS-aided ISAC systems is considered in \cite{LiuR3}. The multi-user sum-rate maximization under radar SNR constraint or CRB constraint is investigated in \cite{LiuR4}.
In \cite{ZhangH}, the authors explore the performance trade-off maximization between communication data rate and sensing mutual information (MI). Robust beamforming design for RIS-aided ISAC systems under imperfect angle knowledge and channel state information (CSI) is studied in \cite{LuanM}. In \cite{WangX}, \cite{WangX2}, RIS is deployed in ISAC systems to mitigate MUI while guaranteeing the radar sensing beampattern and CRB constraint, respectively. Moreover, different from previous studies on narrowband scenarios, recent works \cite{WeiT}, \cite{WeiT1} investigate wideband RIS-aided ISAC systems in conjunction with orthogonal frequency division multiplexing (OFDM) technology.

While numerous studies on RIS have demonstrated its advantages in both communication and ISAC systems, its fatal defect, i.e., the ``multiplicative fading'' effect, has also been exposed. The equivalent path-loss of the RIS-introduced reflection link is a product of the path-losses of the transmitter-RIS link and the RIS-receiver link, thus this phenomenon can be mathematically described as the signal from the transmitter to the receiver via RIS undergoing multiplicative fading. Consequently, in the case of a strong direct link or when the receiver is not close enough to RIS, the performance improvement from the reflection link provided by the passive RIS is marginal. Active RIS is an emerging technology proposed to effectively mitigate the multiplicative fading issue existing in conventional passive RIS-aided systems \cite{ZhangZ}, \cite{LongR}. By integrating reflection-type amplifiers into existing passive electromagnetic components, active RIS is capable of not only reflecting the incident signal with the desired phase-shift, but also amplifying the reflected signal, thereby efficiently compensating the path-loss.
Recent research works \cite{ZhangZ}-\cite{ZhuQ0} have confirmed the advantages of active RIS and further investigated its various communication application scenarios.
Although the active RIS has potential drawbacks and limitations in that it is more costly and has more power consumption than the passive one due to the integration of active amplifiers, a satisfactory balance between performance and cost/power consumption can be achieved by appropriately selecting the number of reflection elements. In general, active RIS is increasingly becoming one of the key enablers to realize more spectrum- and energy-efficient wireless communications.

Owing to its significant advantage in overcoming severe path-loss, active RIS intrinsically possesses immense potential in ISAC systems. The reason for this is that the receivers in practical ISAC systems are typically less sensitive than traditional radar receivers, primarily due to cost considerations related to hardware.
Hence, the reception of weak echo signals by the low-sensitivity ISAC receivers results in unsatisfactory target detection/parameter estimation performance. Active RIS has become a prospective solution for ISAC systems to address the above issues and enhance both radar echo signal quality and communication performance by situationally manipulating the wireless propagations and amplifying the signals. There are several studies intended to explore the application of active RIS in ISAC systems. The authors in \cite{AA} propose to utilize an active RIS to improve the achievable communication secrecy rate while taking into account the worst radar detection SNR. Moreover, an active RIS-aided ISAC system in the scenario of cloud radio access network (C-RAN) is investigated in \cite{ZhangY}. Our recent work \cite{ZhuQ} employs active RIS to overcome the blockage issue by introducing an additional virtual LoS link between the base station (BS) and the target. Both transmit/receive and reflection beamformings are jointly designed to maximize the radar SNR while guaranteeing pre-defined SINRs for communication users.
While existing works on active RIS-empowered ISAC systems focus on target detection function, target parameter estimation is also an important task in radar sensing and should be further explored.

Motivated by the aforementioned discussions, we investigate the deployment of active RIS in ISAC systems in this paper, with an emphasis on the parameter estimation function for the radar sensing component. Specifically, we consider an ISAC system where BS communicates with multiple users and simultaneously senses a point target blocked by an obstacle. An active RIS is employed to support both communication and sensing functions. Our goal is to jointly design the BS transmit precoding and the active RIS reflection beamforming to optimize the direct-of-arrival (DoA) estimation performance and satisfy the users' quality of service (QoS) demands and the power limitations at the BS and the active RIS. The main novelties and contributions of this paper are summarized as follows.
\begin{itemize}
\item Firstly, we introduce active RIS in ISAC systems to enhance the radar parameter estimation performance while guaranteeing the quality of multi-user communications. We formulate signal models for the reception at both the communication users and the BS, from which we derive performance metrics for communication and radar sensing, respectively. More specifically, the CRB for the target DoA estimation in this considered active RIS-empowered ISAC system is meticulously derived for the first time, which is quite different from the CRB for passive RIS-assisted ISAC systems.
While the CRB of DoA estimation is utilized to evaluate the sensing performance of target DoA estimation, the SINR of each user is employed to assess the communication performance.
\item Then, we formulate the joint transmit precoding and active RIS beamforming design problem that aims at minimizing the CRB for target DoA estimation, subject to communication users' SINR requirements, power limitations at the BS and the active RIS, and amplitude constraints of the active RIS reflection coefficients. In an effort to tackle the intricate joint design challenge due to the introduction of active RIS, we develop an effective algorithm that leverages alternating optimization, semidefinite relaxation (SDR), and majorization-minimization (MM) methods.
\item Finally, we provide extensive simulation results to verify the advantages of the proposed active RIS-empowered ISAC scheme and the effectiveness of the developed joint design algorithm. Notably, active RIS can offer over 30dB CRB reduction compared to passive RIS-assisted ISAC systems, thereby achieving substantial sensing performance improvement.
\end{itemize}

\textit{Notations}: $a$ is a scalar, $\mathbf{a}$ is a column vector, and $\mathbf{A}$ is a matrix, respectively. $\mathbf{A}^{T}$, $\mathbf{A}^{*}$, $\mathbf{A}^{H}$ and $\mathbf{A}^{-1}$ denote the transpose, conjugate, Hermitian (conjugate transpose) and inverse operations, respectively. $|a|$, $\|\mathbf{a}\|_2$ and $\|\mathbf{A}\|_F$ denote the magnitude of a scalar $a$, the norm of a vector $\mathbf{a}$ and the Frobenius norm of matrix $\mathbf{A}$. $\text{diag}\{\mathbf{a}\}$ is a diagonal matrix whose diagonal elements are extracted from vector $\mathbf{a}$. $\text{rank}\{\mathbf{A}\}$ is the rank of the matrix $\mathbf{A}$, $\text{Tr}\{\mathbf{A}\}$ is the trace of the matrix $\mathbf{A}$ and $\text{vec}\{\mathbf{A}\}$ denotes vectorization of the matrix $\mathbf{A}$. $\mathbf{I}_{N}$ is an identity matrix of $N$ dimension and $\mathbf{0}$ refers to an all-zeros vector. $\otimes$ is the Kronecker product. $\mathbb{C}$ represents the set of complex numbers. $\Re\{\cdot\}$ and $\Im\{\cdot\}$ extract the real part and imaginary part of a complex number, respectively.


\section{System Model and Problem Formulation}

\begin{figure}[t]
\centering
  \includegraphics[width = 3.4 in]{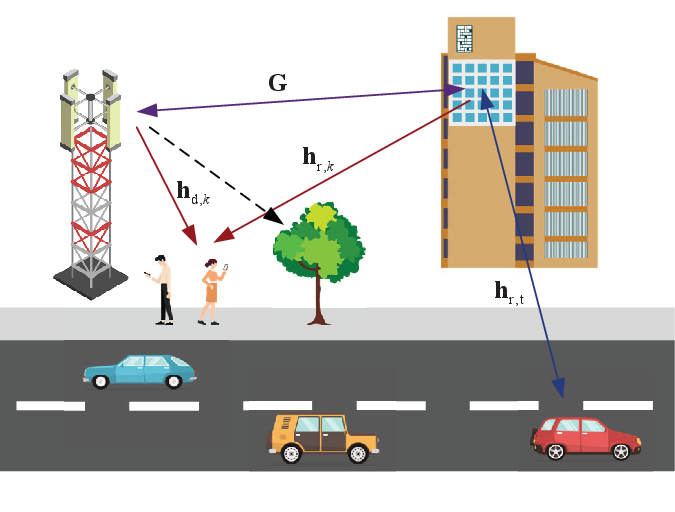}
  \caption{An active RIS-empowered ISAC system.}
  \label{fig:system_model}
\end{figure}

We consider an active RIS-empowered ISAC system as depicted in Fig. \ref{fig:system_model}, in which a dual-functional BS simultaneously performs multi-user communication and co-located radar sensing functions. Particularly, the BS equipped with $N_\mathrm{t}$ transmit antennas and $N_\mathrm{r}$ receive antennas communicates with $K$ single-antenna users and senses one potential target with the aid of an $M$-element active RIS. The target is located in the blind zone of the BS, where the direct BS-target link is blocked by obstacles. For simplicity, we adopt the assumption that $N_\mathrm{t} = N_\mathrm{r} = N$ in the following. To achieve both satisfactory communication and sensing performance, the transmit dual-functional signal in the $l$-th time slot is composed of precoded communication symbols and radar signals, which can be expressed as
\begin{equation}\label{eq:transmit_signal}
  \mathbf{x}[l] = \mathbf{W}_{\mathrm{c}}\mathbf{s}_{\mathrm{c}}[l]+\mathbf{W}_{\mathrm{r}}\mathbf{s}_{\mathrm{r}}[l] = \mathbf{W}\mathbf{s}[l],
\end{equation}
where $\mathbf{s}_{\mathrm{c}}[l] \in \mathbb{C}^K$ denotes the communication symbols satisfying $\mathbb{E}\{\mathbf{s}_{\mathrm{c}}[l] \mathbf{s}_{\mathrm{c}}[l]^{H}\} =\mathbf{I}_K$ and $\mathbf{s}_{\mathrm{r}}[l] \in \mathbb{C}^N$ denotes the radar signals with $\mathbb{E}\{\mathbf{s}_{\mathrm{r}}[l] \mathbf{s}_{\mathrm{r}}[l]^{H}\} =\mathbf{I}_N$, $\mathbb{E}\{\mathbf{s}_{\mathrm{c}}[l] \mathbf{s}_{\mathrm{r}}[l]^{H}\} =\mathbf{0}$. $\mathbf{W}_{\mathrm{c}} \in \mathbb{C}^{N \times K}$ and $\mathbf{W}_{\mathrm{r}} \in \mathbb{C}^{N \times N}$ represent the communication/radar beamforming matrices, respectively. Furthermore, we define $\mathbf{W} \eqdef [\mathbf{W}_{\mathrm{c}}~\mathbf{W}_{\mathrm{r}}]$ and $\mathbf{s}[l] \eqdef [\mathbf{s}_{\mathrm{c}}[l]^T~\mathbf{s}_{\mathrm{r}}[l]^T]^T$ for brevity.

\subsection{Communication Signal Model}
The received signal at the $k$-th communication user is presented as
\begin{equation}\label{eq:received_signal_user_k}
  y_k[l] = (\mathbf{h}_{\mathrm{d}, k}^T + \mathbf{h}_{\mathrm{r}, k}^T\mathbf{\Phi}\mathbf{G})\mathbf{x}[l] + \mathbf{h}_{\mathrm{r}, k}^T\mathbf{\Phi}\mathbf{z}_0[l] + n_k[l],
\end{equation}
where $\mathbf{h}_{\mathrm{d}, k} \in \mathbb{C}^N$ represents the channel between the BS and the $k$-th user, $\mathbf{G}\in \mathbb{C}^{M \times N}$ represents the channel between the BS and the active RIS, and $\mathbf{h}_{\mathrm{r}, k} \in \mathbb{C}^M$ represents the channel between the active RIS and the $k$-th user. With advanced channel estimation methods, we assume that the channel is perfectly known.
$\mathbf{\Phi} \in \mathbb{C}^{M \times M}$ denotes the active RIS reflection beamforming matrix with $\mathbf{\Phi} \eqdef \text{diag}\{\boldsymbol{\phi}\}$, in which $\boldsymbol{\phi} \eqdef [ \phi_1,\ldots, \phi_M]^T \in \mathbb{C}^M$ is the reflection coefficient vector. Unlike traditional passive RIS which only has the capacity to tune the phase-shift of the incident signal, active RIS can further amplify its amplitude thanks to the integration of additional amplifiers into each electromagnetic element.
It is assumed that all active RIS reflection elements are realized by the same type of amplifier, namely, they have the same amplification capacity.
Accordingly, we can formulate the reflection coefficient of the active RIS as $\phi_m \eqdef a_m e^{\jmath\varphi_m}, ~\forall m$, and the amplitude $a_m$ can be continuously adjusted within the interval $a_m \in (0, a_{\max}]$, in which $a_{\max} \geq 1$ is the maximum signal amplification provided by the amplifier.
In addition, $\mathbf{z}_0[l] \sim \mathcal{C} \mathcal{N} (\mathbf{0}, \sigma^{2}_\mathrm{z}\mathbf{I}_{M})$ and $n_k[l] \sim \mathcal{C} \mathcal{N} (0, \sigma_k^{2})$ denote additive white Gaussian noise (AWGN) at the active RIS and the $k$-th user, respectively.

For multi-user communications, the widely-used SINR is adopted as the performance metric. Based on the received signal expression in (\ref{eq:received_signal_user_k}), the SINR of the $k$-th user can be calculated as
\begin{equation}\label{eq:SINR_user_k}
  \text{SINR}_{k}=\frac{|\mathbf{h}_{k}^{T} \mathbf{w}_{k}|^{2}}{\sum_{i \neq k}^{K+N}|\mathbf{h}_{k}^{T} \mathbf{w}_{i}|^{2}+\|\mathbf{h}_{\mathrm{r}, k}^{T} \mathbf{\Phi}\|^{2}_2 \sigma_{\mathrm{z}}^{2}+\sigma_k^{2}},
\end{equation}
where we define $\mathbf{h}_{k}^{T} \triangleq \mathbf{h}_{\mathrm{d}, k}^{T} + \mathbf{h}_{\mathrm{r}, k}^{T} \mathbf{\Phi} \mathbf{G}$ as the equivalent compound channel between the BS and the $k$-th user and $\mathbf{w}_i$ as the $i$-th column of $\mathbf{W}$.

\subsection{Radar Signal Model}
Since the direct BS-target link is blocked by obstacles, the transmitted signal can only reach the target through the active RIS-assisted reflected link and return via the same path. Therefore, the received echo signal at the BS is denoted as
\begin{equation}\label{eq:received_signal_BS}
  \mathbf{y}_{\mathrm{r}}[l] = \mathbf{G}^T\mathbf{\Phi}\Big(\mathbf{h}_{\mathrm{r}, \mathrm{t}} {\alpha} \mathbf{h}_{\mathrm{r}, \mathrm{t}}^T\mathbf{\Phi}(\mathbf{G}\mathbf{x}[l] + \mathbf{z}_{0}[l]) + \mathbf{z}_1[l] \Big)+ \mathbf{n}_{\mathrm{r}}[l],
\end{equation}
where ${\alpha} \sim \mathcal{CN} (0,\sigma_\mathrm{t}^2)$ is the target's radar cross section (RCS) and $\mathbf{h}_{\mathrm{r}, \mathrm{t}} \in \mathbb{C}^M$ is the LoS channel from the active RIS to the target. Specifically, $\mathbf{h}_{\mathrm{r}, \mathrm{t}} \eqdef \alpha_{\mathrm{r,t}}\mathbf{a}_M(\theta)$, where $\alpha_{\mathrm{r,t}}$ represents the path-loss and $\mathbf{a}_M(\theta) \eqdef [1, e^{\jmath\pi\sin\theta}, \ldots, e^{\jmath(M-1)\pi\sin\theta}]^T$ represents the steering vector with $\theta$ being the DoA of the target with respect to the active RIS.
We assume that the target is potentially located at a certain angle/range detection cell. As a result, the DoA with respect to the active RIS and the LoS link of the RIS-target channel are considered known for the purpose of calculating the target estimation CRB.
Moreover, $\mathbf{z}_1[l] \sim \mathcal{C} \mathcal{N} (\mathbf{0}, \sigma^{2}_\mathrm{z}\mathbf{I}_{M})$ and $\mathbf{n}_\mathrm{r}[l] \sim \mathcal{C} \mathcal{N} (\mathbf{0}, \sigma_\mathrm{r}^{2}\mathbf{I}_{N})$ are AWGN at the active RIS and the BS, respectively. Since the noise signal $\mathbf{z}_0[l]$ undergoes multiple attenuations through the RIS-target-RIS-BS link, when the echo signal reaches the BS, its power is substantially smaller than that of other signals, typically of the order of $10^6$, thereby can be ignored. Then, the received signal $\mathbf{y}_{\mathrm{r}}[l]$ can be further approximated as
\begin{equation}\label{eq:received_signal_BS2}
  \mathbf{y}_{\mathrm{r}}[l] \approx {\alpha}\mathbf{G}^T\mathbf{\Phi}\mathbf{h}_{\mathrm{r}, \mathrm{t}} \mathbf{h}_{\mathrm{r}, \mathrm{t}}^T\mathbf{\Phi}\mathbf{G}\mathbf{W}\mathbf{s}[l] + \mathbf{G}^T\mathbf{\Phi}\mathbf{z}_1[l] + \mathbf{n}_{\mathrm{r}}[l].
\end{equation}
By stacking $L$ samples, we can express the combined received signals as
\begin{equation}\label{eq:received_signal_BS_matrix}
  \mathbf{Y}_{\mathrm{r}} = \alpha\mathbf{Q}\mathbf{W}\mathbf{S} + \mathbf{G}^T\mathbf{\Phi}\mathbf{Z}_1 + \mathbf{N}_{\mathrm{r}},
\end{equation}
where we define $\mathbf{Q} \eqdef \mathbf{G}^T\mathbf{\Phi}\mathbf{h}_{\mathrm{r}, \mathrm{t}} \mathbf{h}_{\mathrm{r}, \mathrm{t}}^T\mathbf{\Phi}\mathbf{G}$ and the symbol/noise matrices as $\mathbf{S} \eqdef [\mathbf{s}[1], \ldots, \mathbf{s}[L]]$, $\mathbf{Z}_1 \eqdef [\mathbf{z}_1[1], \ldots, \mathbf{z}_1[L]]$ and $\mathbf{N}_{\mathrm{r}} \eqdef [\mathbf{n}_{\mathrm{r}}[1], \ldots, \mathbf{n}_{\mathrm{r}}[L]]$, respectively. As the dual-functional BS executes the co-located radar sensing function, it has full knowledge of the transmit signal. Therefore, the entire first term in (\ref{eq:received_signal_BS_matrix}), including both the communication and sensing components, is regarded as a useful signal for estimating target parameters, while the second and third terms are considered as noise signals. The vectorized signal is further denoted as
\begin{equation}\label{eq:received_signal_BS_vector}
  \widetilde{\mathbf{y}} = \text{vec}\{\mathbf{Y}_{\mathrm{r}}\} = \mathbf{y} + \mathbf{n},
\end{equation}
in which we apply the definitions of $\mathbf{y} \eqdef \alpha \text{vec}\{\mathbf{Q}\mathbf{W}\mathbf{S}\}$ and $\mathbf{n} \eqdef \text{vec}\{\mathbf{G}^T\mathbf{\Phi}\mathbf{Z}_1 + \mathbf{N}_{\mathrm{r}}\}$.

From the perspective of radar sensing, we focus on parameter estimation performance in terms of CRB. It offers a lower-bound on any unbiased estimator and is an extensively utilized radar metric. Let $\bolds{\xi} \eqdef [\theta, {\bolds{\alpha}}^T]^T$ denote the target parameters to be estimated with $\bolds{\alpha} \eqdef [\Re\{\alpha\}, \Im\{\alpha\}]^T$.
The CRB matrix is the inverse of the Fisher information matrix (FIM) for estimating $\bolds{\xi}$ \cite{SMK}.
In order to facilitate the derivation of CRB, we let $\mathbf{M} \in \mathbb{C}^{3\times 3}$ represent the FIM. As presented in \cite{SMK}, based on the complex observation $\widetilde{\mathbf{y}} \sim \mathcal{CN}(\mathbf{y}, \mathbf{R}_\mathrm{n})$ with $\mathbf{R}_\mathrm{n} \eqdef \mathbf{I}_L \otimes (\sigma_\mathrm{z}^2\mathbf{G}^T\mathbf{\Phi}\mathbf{\Phi}^H\mathbf{G}^* + \sigma_\mathrm{r}^2\mathbf{I}_{N})$, each element of $\bf{M}$ for estimating the unknown parameters $\boldsymbol{\xi}$ can be calculated by
\begin{equation}\label{M}
\mathbf{M}(i,j) = \text{Tr}\{\mathbf{R}_\mathrm{n}^{-1}\frac{\partial \mathbf{R}_\mathrm{n}}{\partial \bolds{\xi}_i} \mathbf{R}^{-1}_\mathrm{n} \frac{\partial \mathbf{R}_\mathrm{n}}{\partial \bolds{\xi}_j}\} + 2\Re\{\frac{\partial {\mathbf{y}}^H}{\partial \bolds{\xi}_i} \mathbf{R}^{-1}_\mathrm{n} \frac{\partial {\mathbf{y}}}{\partial \bolds{\xi}_j}\}.
\end{equation}
According to the definition of $\mathbf{y} \triangleq \alpha \text{vec}\{\mathbf{Q}\mathbf{W}\mathbf{S}\}$, the derivatives of $\mathbf{y}$ with respect to each parameter can be obtained as\begin{subequations}\label{y_xi}
\begin{align}
  \frac{\partial {\mathbf{y}}}{\partial \theta} &= \alpha \text{vec}\{\dot{\mathbf{Q}}\mathbf{W}\mathbf{S}\},\\
  \frac{\partial {\mathbf{y}}}{\partial {\boldsymbol{\alpha}}} &= [1,\jmath] \otimes \text{vec}\{{\mathbf{Q}}\mathbf{W}\mathbf{S}\},
\end{align}
\end{subequations}
where $\dot{\mathbf{Q}} \eqdef \frac{\partial {\mathbf{Q}}}{\partial \theta}$ denotes the partial derivative of $\mathbf{Q}$ in terms of $\theta$ \cite{IB}. Recalling $\mathbf{Q} \eqdef \mathbf{G}^T\mathbf{\Phi}\mathbf{h}_{\mathrm{r}, \mathrm{t}} \mathbf{h}_{\mathrm{r}, \mathrm{t}}^T\mathbf{\Phi}\mathbf{G}$, $\mathbf{h}_{\mathrm{r}, \mathrm{t}} \eqdef \alpha_{\mathrm{r,t}}\mathbf{a}_M(\theta)$, and defining $\mathbf{q} \eqdef \mathbf{G}^T\mathbf{\Phi}\mathbf{a}_M(\theta)$, $\mathbf{Q}$ can be re-expressed as $\mathbf{Q} = \alpha_\mathrm{r,t}^2\mathbf{q}\mathbf{q}^T$. Thus, the partial derivative $\dot{\mathbf{Q}}$ is written as
\begin{equation}\label{eq:Q_der}
\begin{aligned}
  \dot{\mathbf{Q}} &= \alpha_\mathrm{r,t}^2(\dot{\mathbf{q}}{\mathbf{q}}^T + {\mathbf{q}}\dot{\mathbf{q}}^T)\\
  &= c_0 \mathbf{G}^T {\mathbf{A}}(\mathbf{L}\bolds{\phi}\bolds{\phi}^T + \bolds{\phi}\bolds{\phi}^T\mathbf{L}){\mathbf{A}}\mathbf{G},
\end{aligned}
\end{equation}
where $\dot{\mathbf{q}}$ is the partial derivative of $\mathbf{q}$ with respect to $\theta$, which is derived as $\dot{\mathbf{q}} \eqdef \frac{\partial {\mathbf{q}}}{\partial \theta} = \jmath \pi \cos\theta\mathbf{G}^T\mathbf{\Phi} \mathbf{L} \mathbf{a}_M(\theta) = \jmath \pi \cos\theta\mathbf{G}^T \mathbf{A}\mathbf{L}\bolds{\phi}$ with $\mathbf{L} \eqdef \text{diag}\{0,1, \cdots, M-1\}$ and ${\mathbf{A}} \eqdef \text{diag}\{\mathbf{a}_M(\theta)\}$. In addition, we define $c_0 \eqdef \alpha_\mathrm{r,t}^2\jmath \pi \cos\theta$ for simplicity.
Moreover, it is obvious that $\mathbf{R}_\mathrm{n} \eqdef \mathbf{I}_L \otimes (\sigma_\mathrm{z}^2\mathbf{G}^T\mathbf{\Phi}\mathbf{\Phi}^H\mathbf{G}^* + \sigma_\mathrm{r}^2\mathbf{I}_{N})$ is irrelevant to $\bolds{\xi}$, that is, $\frac{\partial \mathbf{R}_\mathrm{n}}{\partial \bolds{\xi}_i} = 0, ~\forall i$.
For ease of subsequent handling, we partition $\bf{M}$ into $2 \times 2$ blocks as
\begin{equation}\label{M_matrix}
  \mathbf{M} = \begin{bmatrix}
  \mathbf{M}_{\theta\theta} &  \mathbf{M}_{\theta{\bolds{\alpha}}}\\
   \mathbf{M}_{\theta{\bolds{\alpha}}}^T & \mathbf{M}_{{\bolds{\alpha}}{\bolds{\alpha}}}
  \end{bmatrix}.
\end{equation}
Accordingly, based on the above derivations, $\mathbf{M}_{\theta\theta}$ can be calculated as\begin{subequations}\label{M11}
\begin{align}
  &\mathbf{M}_{\theta\theta} =  2\Re\{\frac{\partial {\mathbf{y}}^H}{\partial {\theta}} \mathbf{R}^{-1}_\mathrm{n} \frac{\partial {\mathbf{y}}}{\partial {\theta}}\}\label{M11a}\\
  &~ = 2 \Re\{\alpha^* \text{vec}^H\{\dot{\mathbf{Q}}\mathbf{W}\mathbf{S}\} (\mathbf{I}_L \otimes \widetilde{\mathbf{R}}_\mathrm{n}^{-1}) \alpha \text{vec}\{\dot{\mathbf{Q}}\mathbf{W}\mathbf{S}\}\} \label{M11b}\\
  &~ = 2 |\alpha|^2 \text{Tr}\{\dot{\mathbf{Q}}\mathbf{W}\mathbf{S}\mathbf{S}^H\mathbf{W}^H\dot{\mathbf{Q}}^H\widetilde{\mathbf{R}}_\mathrm{n}^{-1}\}\label{M11c} \\
  &~ = 2 L |\alpha|^2 \text{Tr}\{\dot{\mathbf{Q}}\mathbf{W}\mathbf{W}^H\dot{\mathbf{Q}}^H\widetilde{\mathbf{R}}_\mathrm{n}^{-1}\}\label{M11d},
\end{align}
\end{subequations}
where (\ref{M11a})-(\ref{M11b}) holds since we re-denote $\mathbf{R}_\mathrm{n} \eqdef \mathbf{I}_L \otimes \widetilde{\mathbf{R}}_\mathrm{n}$ with $\widetilde{\mathbf{R}}_\mathrm{n} \eqdef \sigma_\mathrm{z}^2\mathbf{G}^T\mathbf{\Phi}\mathbf{\Phi}^H\mathbf{G}^* + \sigma_\mathrm{r}^2\mathbf{I}_{N}$ and apply the property $\mathbf{R}_\mathrm{n}^{-1} = \mathbf{I}_L \otimes \widetilde{\mathbf{R}}_\mathrm{n}^{-1}$. The transformations $\text{Tr}\{\mathbf{A}\mathbf{B}\mathbf{C}\mathbf{D}\} = \text{vec}^H\{\mathbf{D}^H\}(\mathbf{C}^T \otimes \mathbf{A})\text{vec}\{\mathbf{B}\}$ and $\text{Tr}\{\mathbf{A}\mathbf{B}\} = \text{Tr}\{\mathbf{B}\mathbf{A}\}$ are utilized to support the conversion in (\ref{M11b})-(\ref{M11c}) \cite{SongX2}. Moreover, it is noted that due to the facts that $\mathbb{E}\{\mathbf{S}\mathbf{S}^H\} = L\mathbf{I}_{N+K} $ and sufficient samples are usually collected for parameter estimation, we assume that $\mathbf{S}\mathbf{S}^H = L\mathbf{I}_{N+K}$ in (\ref{M11c})-(\ref{M11d}).
Similarly, $\mathbf{M}_{\theta{\bolds{\alpha}}}$ and $\mathbf{M}_{\bolds{\alpha}{\bolds{\alpha}}}$ are given by\begin{subequations}\label{M12}
\allowdisplaybreaks[4]
\begin{align}
    &\mathbf{M}_{\theta{\bolds{\alpha}}} =  2\Re\{\frac{\partial {\mathbf{y}}^H}{\partial {\theta}} \mathbf{R}^{-1}_\mathrm{n} \frac{\partial {\mathbf{y}}}{\partial {{\bolds{\alpha}}}}\}\\
  &~ = 2 \Re\{\alpha^* \text{vec}^H\{\dot{\mathbf{Q}}\mathbf{W}\mathbf{S}\} \mathbf{R}^{-1}_\mathrm{n} ([1,\jmath] \otimes \text{vec}\{{\mathbf{Q}}\mathbf{W}\mathbf{S}\})\}\\
  &~ = 2 \Re\{\alpha^* \text{vec}^H\{\dot{\mathbf{Q}}\mathbf{W}\mathbf{S}\} \mathbf{R}^{-1}_\mathrm{n}\text{vec}\{{\mathbf{Q}}\mathbf{W}\mathbf{S}\} [1,\jmath]\}\\
  &~ = 2 L \Re\{\alpha^*\text{Tr}\{{\mathbf{Q}}\mathbf{W}\mathbf{W}^H\dot{\mathbf{Q}}^H\widetilde{\mathbf{R}}_\mathrm{n}^{-1}\}[1,\jmath]\},
\end{align}
\end{subequations}

\vspace{-0.4cm}

\begin{subequations}\label{M22}
\allowdisplaybreaks[4]
\begin{align}
  \hspace{0.025cm}&\mathbf{M}_{{\bolds{\alpha}}{\bolds{\alpha}}} =  2\Re\{\frac{\partial {\mathbf{y}}^H}{\partial {{\bolds{\alpha}}}} \mathbf{R}^{-1}_\mathrm{n} \frac{\partial {\mathbf{y}}}{\partial {{\bolds{\alpha}}}}\}\\
  &~ = 2 \Re\{([1,\jmath] \otimes \text{vec}\{{\mathbf{Q}}\mathbf{W}\mathbf{S}\})^H {\mathbf{R}}_\mathrm{n}^{-1}\\
   &~\quad \times ([1,\jmath] \otimes \text{vec}\{{\mathbf{Q}}\mathbf{W}\mathbf{S}\})\} \notag\\
  &~ = 2 \Re\{([1,\jmath]^H[1,\jmath]) \otimes(\text{vec}^H\{{\mathbf{Q}}\mathbf{W}\mathbf{S}\} \\
   &~\quad \times (\mathbf{I}_L \otimes \widetilde{\mathbf{R}}_\mathrm{n}^{-1}) \text{vec}\{{\mathbf{Q}}\mathbf{W}\mathbf{S}\})\}  \notag\\
  &~ = 2 \Re\{([1,\jmath]^H[1,\jmath]) \otimes \text{Tr}\{{\mathbf{Q}}\mathbf{W}\mathbf{S}\mathbf{S}^H\mathbf{W}^H{\mathbf{Q}}^H\widetilde{\mathbf{R}}_\mathrm{n}^{-1}\}\} \\
  &~ = 2 L \text{Tr}\{{\mathbf{Q}}\mathbf{W}\mathbf{W}^H{\mathbf{Q}}^H\widetilde{\mathbf{R}}_\mathrm{n}^{-1}\}\mathbf{I}_2.
\end{align}
\end{subequations}

After the complete derivation of FIM, the CRB for the $i$-th parameter to be estimated (i.e., the $i$-th element of $\bolds{\xi}$) can be found as the $(i,i)$-th element of the inverse of $\mathbf{M}$ \cite{SMK}.
In the considered scenario, the dual-functional BS performs a radar function to sense the target's direction. Thus we are more interested in estimating the target's DoA\footnote{Considering factors such as application scenarios, resource limitations, and algorithm complexity, in this initial work, we focus exclusively on CRB optimization for DoA estimation. The CRB optimization for RCS/range estimation will be investigated in our future work.}, i.e. the parameter $\theta$ which is the first element of $\bolds{\xi}$.
Therefore, substituting the results in (\ref{M11}), (\ref{M12}) and (\ref{M22}) and applying the inverse matrix definition of a partitioned matrix, the CRB for estimating the target's DoA $\theta$ can be denoted as \cite{SongX2}
\begin{equation}\label{CRB}
	\allowdisplaybreaks[4]
\begin{aligned}
  &\text{CRB}_{\theta} = [\mathbf{M}^{-1}]_{1,1} = [\mathbf{M}_{\theta\theta} - \mathbf{M}_{\theta{\bolds{\alpha}}}\mathbf{M}_{{\bolds{\alpha}}{\bolds{\alpha}}}^{-1}\mathbf{M}_{\theta{\bolds{\alpha}}}^T]^{-1}\\
  = &\frac{1}{2 L |\alpha|^2 \Big(\text{Tr}\{\dot{\mathbf{Q}}\mathbf{W}\mathbf{W}^H\dot{\mathbf{Q}}^H\widetilde{\mathbf{R}}_\mathrm{n}^{-1}\}
  - \frac{|\text{Tr}\{{\mathbf{Q}}\mathbf{W}\mathbf{W}^H\dot{\mathbf{Q}}^H\widetilde{\mathbf{R}}_\mathrm{n}^{-1}\}|^2}
  {\text{Tr}\{{\mathbf{Q}}\mathbf{W}\mathbf{W}^H{\mathbf{Q}}^H\widetilde{\mathbf{R}}_\mathrm{n}^{-1}\}}\Big)}.
\end{aligned}
\end{equation}

\subsection{Problem Formulation}
In this paper, we aim at minimizing the CRB for estimating the target's DoA $\theta$ by jointly designing transmit precoding $\mathbf{W}$ and active RIS reflection beamforming $\bolds{\phi}$ to improve the radar parameter estimation performance while assuring the QoS requirements of communication users. Consequently, the optimization problem can be formulated as
\begin{subequations}\label{pr:original_problem_crb}
\begin{align}
\min_{\mathbf{W}, \bolds{\phi}} ~~&\text{CRB}_{\theta} \label{original_problema}\\
\text{s.t.} ~~~& \|\mathbf{W}\|^2_{F} \leq P_{\mathrm{BS}}, \label{original_problemb}\\
& \mathcal{P}(\mathbf{W},\bolds{\phi}) \leq P_{\mathrm{RIS}}, \label{original_problemc}\\
& \text{SINR}_{k} \geq \gamma_{k},~\forall k, \label{original_problemd}\\
& a_m \leq a_{\max},~\forall m, \label{original_probleme}
\end{align}
\end{subequations}
where $\mathcal{P}(\mathbf{W},\bolds{\phi})$ is the power consumption at the active RIS, denoted as
\begin{equation}\label{eq:P_ris}
\begin{aligned}
  \mathcal{P}(\mathbf{W},\bolds{\phi}) &= \|\mathbf{\Phi}\mathbf{G}\mathbf{W}\|^2_F + \sigma_\mathrm{t}^2\|\mathbf{\Phi}\mathbf{h}_{\mathrm{r}, \mathrm{t}}\mathbf{h}_{\mathrm{r}, \mathrm{t}}^T\mathbf{\Phi}\mathbf{G}\mathbf{W}\|^2_F \\ & \quad +\sigma_\mathrm{t}^2\sigma_\mathrm{z}^2\|\mathbf{\Phi}\mathbf{h}_{\mathrm{r}, \mathrm{t}}\mathbf{h}_{\mathrm{r}, \mathrm{t}}^T\mathbf{\Phi}\|^2_{F} + 2\sigma_\mathrm{z}^2\|\mathbf{\Phi}\|^2_F.
\end{aligned}
\end{equation}
Moreover, (\ref{original_problemb}) and (\ref{original_problemc}) represent the power constraints, where $P_{\mathrm{BS}}$ and $P_{\mathrm{RIS}}$ are the power budgets at the BS/active RIS, respectively. In general, the transmit power budget of the BS is greater than the amplification power budget of the active RIS, varying between $20\sim50$ dBm and $0\sim10$ dBm, respectively. The specific values of $P_\mathrm{BS}$ and $P_\mathrm{RIS}$ for simulation studies will be provided in Section \ref{4}. (\ref{original_problemd}) guarantees the worst-case user's communication SINR $\gamma_k$ and (\ref{original_probleme}) is the maximum amplitude constraint of the active RIS. The complicated expression with fractional terms and the coupling of optimization variables in both objective function and constraints make problem (\ref{pr:original_problem_crb}) non-convex and highly challenging to solve. To overcome these difficulties, we propose to alternatively optimize the sub-problems on each variable.

\section{Joint Transmit Precoding and RIS Reflection Beamforming Design}
\subsection{Objective Transformation}
According to the expression of $\text{CRB}_{\theta}$ in (\ref{CRB}), minimizing $\text{CRB}_{\theta}$ is equivalent to maximizing its denominator, and thereby we can reasonably transform the original objective function in (\ref{pr:original_problem_crb}) into
\begin{equation}\label{objective_function_g}
\begin{aligned}
\max_{\mathbf{W}, \bolds{\phi}} ~~ g(\mathbf{W}, \bolds{\phi}) &\eqdef \text{Tr}\{\dot{\mathbf{Q}}\mathbf{W}\mathbf{W}^H\dot{\mathbf{Q}}^H\widetilde{\mathbf{R}}_\mathrm{n}^{-1}\}
  \\ &\quad- \frac{|\text{Tr}\{{\mathbf{Q}}\mathbf{W}\mathbf{W}^H\dot{\mathbf{Q}}^H\widetilde{\mathbf{R}}_\mathrm{n}^{-1}\}|^2}
  {\text{Tr}\{{\mathbf{Q}}\mathbf{W}\mathbf{W}^H{\mathbf{Q}}^H\widetilde{\mathbf{R}}_\mathrm{n}^{-1}\}}.
\end{aligned}
\end{equation}
In the following, we alternately design the transmit precoding $\mathbf{W}$ and the active RIS beamforming $\bolds{\phi}$ to maximize the objective function $g(\mathbf{W}, \bolds{\phi})$ for improving CRB performance.

\vspace{-0.2cm}
\subsection{Transmit Precoding $\mathbf{W}$ Design}
With fixed active RIS reflection beamforming vector $\bolds{\phi}$, the sub-problem for optimizing $\mathbf{W}$ can be formulated as
\begin{subequations}\label{pr:sub_problem_w1}
\begin{align}
\max_{\mathbf{W}} ~~&g(\mathbf{W}) \\
\text{s.t.} ~~~& \sum\nolimits_{i = 1}^{K+N} \|\mathbf{w}_i\|_2^2 \leq P_{\mathrm{BS}},\\
& \sum\nolimits_{i = 1}^{K+N}\text{Tr}\{\mathbf{w}_i\mathbf{w}_i^H\mathbf{E}\} \leq P_{\mathrm{RIS}} - c_\mathrm{r},\\
& (1+\gamma_{k}^{-1})\mathbf{h}_{k}^{T} \mathbf{w}_{k}\mathbf{w}_{k}^H\mathbf{h}_{k}^{*} \\
&\geq  \sum\nolimits_{i = 1}^{K+N}\mathbf{h}_{k}^{T} \mathbf{w}_{i}\mathbf{w}_{i}^H\mathbf{h}_{k}^{*} + c_\mathrm{s},~\forall k \notag,
\end{align}
\end{subequations}
where for clarity we define
\begin{subequations}\label{eq:trans_w}
\begin{align}
\mathbf{E} &\triangleq \mathbf{G}^H\mathbf{\Phi}^H\mathbf{\Phi}\mathbf{G} +  \sigma_\mathrm{t}^2 \mathbf{G}^H\mathbf{\Phi}^H\mathbf{h}_{\mathrm{r}, \mathrm{t}}^*\mathbf{h}_{\mathrm{r}, \mathrm{t}}^H\mathbf{\Phi}^H\mathbf{\Phi}\mathbf{h}_{\mathrm{r}, \mathrm{t}}\mathbf{h}_{\mathrm{r}, \mathrm{t}}^T\mathbf{\Phi}\mathbf{G},\\
c_\mathrm{r} & \eqdef \sigma_\mathrm{t}^2\sigma_\mathrm{z}^2\|\mathbf{\Phi}\mathbf{h}_{\mathrm{r}, \mathrm{t}}\mathbf{h}_{\mathrm{r}, \mathrm{t}}^T\mathbf{\Phi}\|^2_{F} + 2\sigma_\mathrm{z}^2\|\mathbf{\Phi}\|^2_F,\\
c_\mathrm{s} &\eqdef \|\mathbf{h}_{\mathrm{r}, k}^{T} \mathbf{\Phi}\|^{2}_2 \sigma_\mathrm{z}^{2}+\sigma_k^{2}.
\end{align}
\end{subequations}
In particular, $g(\mathbf{W})$ is a complicated expression with fractional term and higher-order term with respect to $\mathbf{W}$. In order to address the above issues, we propose to introduce a lower-bound for $g(\mathbf{W})$ rather than directly optimizing it, and then invoke the Schur complement and SDR method to effectively solve the sub-problem.

Specifically, we introduce an auxiliary variable $t_\mathrm{w}$ and favorably represent problem (\ref{pr:sub_problem_w1}) as
\begin{subequations}\label{pr:sub_problem_w2}
\begin{align}
\max_{\mathbf{W},t_\mathrm{w}} ~~&t_\mathrm{w}\\
\text{s.t.} ~~~&g(\mathbf{W}) \geq t_\mathrm{w}, \label{pr:sub_problem_w2b}\\
& \sum\nolimits_{i = 1}^{K+N} \|\mathbf{w}_i\|_2^2 \leq P_{\mathrm{BS}},\\
& \sum\nolimits_{i = 1}^{K+N}\text{Tr}\{\mathbf{w}_i\mathbf{w}_i^H\mathbf{E}\} \leq P_{\mathrm{RIS}} - c_\mathrm{r},\\
& (1+\gamma_{k}^{-1})\mathbf{h}_{k}^{T} \mathbf{w}_{k}\mathbf{w}_{k}^H\mathbf{h}_{k}^{*} \label{pr:sub_problem_w2e}\\
&\geq  \sum\nolimits_{i = 1}^{K+N}\mathbf{h}_{k}^{T} \mathbf{w}_{i}\mathbf{w}_{i}^H\mathbf{h}_{k}^{*} + c_\mathrm{s},~\forall k. \notag
\end{align}
\end{subequations}
The constraint in (\ref{pr:sub_problem_w2b}) can be further converted into the below semidefinite form via the Schur complement:

\vspace{-0.3cm}
\begin{small}
\begin{equation}\label{eq:schur}
\begin{bmatrix}
\text{Tr}\{\dot{\mathbf{Q}}\mathbf{W}\mathbf{W}^H\dot{\mathbf{Q}}^H\widetilde{\mathbf{R}}_\mathrm{n}^{-1}\} - t_\mathrm{w} & \text{Tr}\{{\mathbf{Q}}\mathbf{W}\mathbf{W}^H\dot{\mathbf{Q}}^H\widetilde{\mathbf{R}}_\mathrm{n}^{-1}\} \\
\text{Tr}\{\widetilde{\mathbf{R}}_\mathrm{n}^{-1}\dot{\mathbf{Q}}\mathbf{W}\mathbf{W}^H{\mathbf{Q}}^H\} & \text{Tr}\{{\mathbf{Q}}\mathbf{W}\mathbf{W}^H{\mathbf{Q}}^H\widetilde{\mathbf{R}}_\mathrm{n}^{-1}\}
\end{bmatrix} \succeq \mathbf{0}.
\end{equation}
\end{small}

\nid It is easy to see that the constraints (\ref{pr:sub_problem_w2e}) and (\ref{eq:schur}) are still non-convex with respect to variable $\mathbf{W}$ and hard to tackle. Therefore, we propose to convert the optimization variable and further utilize the SDR algorithm for an easier solution. Specifically, by defining
\begin{subequations}\label{eq:Rw}
\begin{align}
  {\mathbf{W}}_i &\eqdef \mathbf{w}_{i}\mathbf{w}_{i}^H, ~ i =  1,\ldots,K+N,\\
   \mathbf{R}_{\mathrm{w}} &\eqdef \sum\nolimits_{i = 1}^{K+N}{\mathbf{W}}_i = \mathbf{W}\mathbf{W}^H,
\end{align}
\end{subequations}
the quadratic terms $\mathbf{w}_{i}\mathbf{w}_{i}^H, ~\forall i$ and $\mathbf{W}\mathbf{W}^H$ are transformed into the forms related to primary variables ${\mathbf{W}}_i, ~\forall i$ and $\mathbf{R}_{\mathrm{w}}$. Meanwhile, these rank-one Hermitian positive semidefinite matrices ${\mathbf{W}}_i, ~\forall i$, and the Hermitian positive semidefinite matrix $\mathbf{R}_{\mathrm{w}}$ should satisfy
\begin{subequations}\label{eq:Rw1}
\begin{align}
  &{\mathbf{W}}_i = {\mathbf{W}}_i^H,~{\mathbf{W}}_i \succeq \mathbf{0}, ~\text{rank}\{{\mathbf{W}}_i\} = 1, ~\forall i,\\
  &\mathbf{R}_{\mathrm{w}} = \mathbf{R}_{\mathrm{w}}^H,~\mathbf{R}_{\mathrm{w}}\succeq \mathbf{0}.
\end{align}
\end{subequations}
It is worth noting that the individual matrices ${\mathbf{W}}_i$, ${i = K+1, \ldots, K+N}$ have no impact on problem (\ref{pr:sub_problem_w2}), and are contained in the matrix $\mathbf{R}_{\mathrm{w}}$. Accordingly, we propose to remove these optimization variables to simplify the transformed optimization problem. Moreover, the rank-one constraint in (\ref{eq:Rw1}a) poses a significant obstacle to finding a straightforward solution, thus, we apply SDR algorithm to relax it. As a result, problem (\ref{pr:sub_problem_w2}) can be converted into
\begin{subequations}\label{pr:sub_problem_w4}
\begin{align}
&\max_{\mathbf{W}_i,i = 1,\ldots, K, \mathbf{R}_{\mathrm{w}},t_\mathrm{w}} ~~t_\mathrm{w}\\
&~\text{s.t.} ~\begin{bmatrix}
\text{Tr}\{\dot{\mathbf{Q}}\mathbf{R}_{\mathrm{w}}\dot{\mathbf{Q}}^H\widetilde{\mathbf{R}}_\mathrm{n}^{-1}\} - t_\mathrm{w} & \text{Tr}\{{\mathbf{Q}}\mathbf{R}_{\mathrm{w}}\dot{\mathbf{Q}}^H\widetilde{\mathbf{R}}_\mathrm{n}^{-1}\} \\
\text{Tr}\{\widetilde{\mathbf{R}}_\mathrm{n}^{-1}\dot{\mathbf{Q}}\mathbf{R}_{\mathrm{w}}{\mathbf{Q}}^H\} & \text{Tr}\{{\mathbf{Q}}\mathbf{R}_{\mathrm{w}}{\mathbf{Q}}^H\widetilde{\mathbf{R}}_\mathrm{n}^{-1}\}
\end{bmatrix} \succeq \mathbf{0},\\
& ~~~~~~\text{Tr}\{\mathbf{R}_{\mathrm{w}}\} \leq P_{\mathrm{BS}},\\
& ~~~~~~\text{Tr}\{\mathbf{R}_{\mathrm{w}}\mathbf{E}\} \leq P_{\mathrm{RIS}} - c_\mathrm{r},\\
& ~~~~~~(1+\gamma_{k}^{-1})\mathbf{h}_{k}^{T} {\mathbf{W}}_k\mathbf{h}_{k}^{*} \geq  \mathbf{h}_{k}^{T} \mathbf{R}_{\mathrm{w}}\mathbf{h}_{k}^{*} + c_\mathrm{s},~\forall k,\\
& ~~~~~~{\mathbf{W}}_i = {\mathbf{W}}_i^H,~{\mathbf{W}}_i \succeq \mathbf{0},~ i = 1, \ldots, K,\\
& ~~~~~~\mathbf{R}_{\mathrm{w}} = \mathbf{R}_{\mathrm{w}}^H,~\mathbf{R}_{\mathrm{w}}\succeq \mathbf{0},\\
& ~~~~~~\mathbf{R}_{\mathrm{w}} - \sum\nolimits_{i = 1}^{K}{\mathbf{W}}_i \succeq \mathbf{0}.
\end{align}
\end{subequations}
Clearly, problem (\ref{pr:sub_problem_w4}) is a convex problem and can be easily solved by standard convex optimization algorithms. After obtaining an optimal solution ${\mathbf{W}}_i, ~i = 1,\ldots, K$ and ${\mathbf{R}}_{\mathrm{w}}$ of (\ref{pr:sub_problem_w4}), the optimal communication beamforming vectors ${\mathbf{w}}_i, ~ i = 1,\ldots, K$ can be recovered as
\begin{equation}\label{eq:W_final}
  {\mathbf{w}}_i = (\mathbf{h}_i^T{\mathbf{R}}_{\mathrm{w}}\mathbf{h}_i^*)^{-1/2}{\mathbf{R}}_{\mathrm{w}}\mathbf{h}_i^*, ~i = 1,\ldots, K.
\end{equation}
The proof of (\ref{eq:W_final}) is given in detail in \cite{LiuX}.

On the other hand, the radar beamforming vectors $\mathbf{w}_i, ~{i = K+1, \ldots, K+N}$ can be calculated by the Cholesky decomposition, i.e.,
\begin{equation}\label{eq:radar_beamforming}
  \mathbf{W}_\mathrm{r}\mathbf{W}_\mathrm{r}^H = \mathbf{R}_{\mathrm{w}} - \sum\nolimits_{i = 1}^{K}{\mathbf{W}}_i,
\end{equation}
where $\mathbf{W}_\mathrm{r} = [\mathbf{w}_{K+1},\ldots, \mathbf{w}_{K+N}]$. Finally, by combining $\mathbf{W} = [\mathbf{w}_1, \cdots, \mathbf{w}_{K+N}]$, the optimal transmit precoding can be achieved.

\subsection{Active RIS Reflection Beamforming $\bolds{\phi}$ Design}
After obtaining transmit precoding $\mathbf{W}$, we focus on the sub-problem of optimizing the active RIS reflection coefficients $\bolds{\phi}$. As demonstrated in (\ref{objective_function_g}),  the quadratic term $\widetilde{\mathbf{R}}_\mathrm{n}$ on $\bolds{\phi}$ exists in $g(\bolds{\phi})$ in the form of an inverse, which is extremely challenging to optimize. To address this difficulty, we propose to first introduce an auxiliary variable $\mathbf{\Psi}$ to take $\widetilde{\mathbf{R}}_\mathrm{n}$ out of the objective function, and then iteratively update the variables $\bolds{\phi}$ and $\mathbf{\Psi}$ until the convergence is achieved.

\vspace{0.3cm}
\subsubsection{Update $\bolds{\phi}$}
Specifically, the optimization with respect to $\bolds{\phi}$ is formulated as
\begin{subequations}\label{pr:objective_phi_1}
	\allowdisplaybreaks[4]
\begin{align}
\max_{\bolds{\phi}} ~~& g(\bolds{\phi}) \label{objective_phi_1a}\\
\text{s.t.} ~~~& \mathcal{P}(\bolds{\phi}) \leq P_{\mathrm{RIS}},\\
& \text{SINR}_{k} \geq \gamma_{k},~\forall k,\\
& a_m \leq a_{\max},~\forall m.
\end{align}
\end{subequations}

In order to promote the algorithm development, we start by reformulating the original problem as an explicit problem with respect to $\bolds{\phi}$, i.e.,
\vspace{0.1cm}
\begin{subequations}\label{pr:phi_10}
\begin{align}
\min_{\boldsymbol{\phi}} ~~ &\frac{\bolds{\xi}_1^H\mathbf{v}\mathbf{v}^H \mathbf{\Xi}_1\mathbf{v}}{\boldsymbol{\phi}^H\mathbf{R}_2\boldsymbol{\phi}}
+\frac{\bolds{\xi}_2^H\mathbf{v}\mathbf{v}^H \mathbf{\Xi}_2\mathbf{v}}{\boldsymbol{\phi}^H\mathbf{R}_1\boldsymbol{\phi}} - \mathbf{v}^H \mathbf{F}\mathbf{v}\\
\text{s.t.} ~~~&\boldsymbol{\phi}^H\mathbf{J}\boldsymbol{\phi}\boldsymbol{\phi}^H\boldsymbol{\phi} + \sigma_\mathrm{t}^2 \sigma_\mathrm{z}^2 \alpha_{\mathrm{r,t}}^4(\boldsymbol{\phi}^H\boldsymbol{\phi})^2\\
 &\quad + \boldsymbol{\phi}^H\mathbf{K}\boldsymbol{\phi} \leq P_{\mathrm{RIS}}, \notag\\
& \boldsymbol{\phi}^H\mathbf{C}_k\boldsymbol{\phi} + \Re\{\mathbf{d}_k^H\boldsymbol{\phi}\} + {c}_{\mathrm{\phi},k}\\
 &\quad - (1+{\gamma_k^{-1}})\boldsymbol{\phi}^H\mathbf{b}_{k,k}^{*}\mathbf{b}_{k,k}^{T}\boldsymbol{\phi} \leq 0,~\forall k, \notag\\
& a_m \leq a_{\max},~\forall m,
\end{align}
\end{subequations}
where $\mathbf{v} \eqdef \text{vec}\{\boldsymbol{\phi}\boldsymbol{\phi}^H\} = \boldsymbol{\phi}^* \otimes \boldsymbol{\phi}$. The proof of equivalence between objective functions (\ref{pr:objective_phi_1}a) and (\ref{pr:phi_10}a) is presented in Appendix \ref{B}. For ease of notation, in (\ref{pr:phi_10}) we define
\begin{subequations}\label{eq:P_phi_trans}
\begin{align}
\mathbf{R}_1 &\eqdef \mathbf{A}^H\mathbf{G}^*\mathbf{W}^*\mathbf{W}^T\mathbf{G}^T\mathbf{A},\\
\mathbf{R}_2 &\eqdef \mathbf{A}^H\mathbf{G}^*\mathbf{\Psi}^{-1}\mathbf{G}^T\mathbf{A},\\
\boldsymbol{\xi}_i &\eqdef \text{vec}\{\mathbf{R}_i^H\}, ~i = 1,2,\\
\mathbf{\Xi}_i &\eqdef \mathbf{L}\mathbf{R}_{{\hat{i}}}^T \otimes \mathbf{L}\mathbf{R}_{{\hat{i}}},~\forall i, ~\hat{i} \neq i,\\
\mathbf{F}_i &\eqdef \mathbf{L}\mathbf{R}_{\hat{i}}^T\mathbf{L} \otimes \mathbf{R}_i,~\forall i, ~\hat{i} \neq i,\\
\mathbf{F} &\eqdef \mathbf{F}_1 + \mathbf{F}_2,\\
  \mathbf{J} &\eqdef \sigma_\mathrm{t}^2 \alpha_{\mathrm{r,t}}^2\text{diag}\{\mathbf{h}_\mathrm{r,t}^*\}\mathbf{G}^*\mathbf{W}^*\mathbf{W}^T\mathbf{G}^T\text{diag}\{\mathbf{h}_\mathrm{r,t}\},\\
  \mathbf{K} &\eqdef \sum\nolimits_{k = 1}^{K+N} \text{diag}\{\mathbf{G}^*\mathbf{w}_k^*\} \text{diag}\{\mathbf{G}\mathbf{w}_k\} + 2\sigma_\mathrm{z}^2 \mathbf{I}_{M},\\
  \mathbf{C}_k &\eqdef \sum\nolimits_{i = 1}^{K+N}\mathbf{b}_{k,i}^{*}\mathbf{b}_{k,i}^{T} + \sigma_\mathrm{z}^2\text{diag}\{\mathbf{h}_{\mathrm{r},k}^H \}\text{diag}\{\mathbf{h}_{\mathrm{r},k} \},\\
  \mathbf{d}_k &\eqdef \sum\nolimits_{i = 1}^{K+N} 2a_{k,i}\mathbf{b}_{k,i}^{*} - 2 (1+{\gamma_k^{-1}})a_{k,k}\mathbf{b}_{k,k}^{*},\\
  {c}_{\mathrm{\phi},k} &\eqdef \sum\nolimits_{i = 1}^{K+N} |a_{k,i}|^2 - (1+{\gamma_k^{-1}})|a_{k,k}|^2 + \sigma_k^2,\\
  a_{k,i} &\eqdef \mathbf{h}_{\mathrm{d},k}^{T}\mathbf{w}_i, ~ \mathbf{b}_{k,i} \eqdef \text{diag}\{\mathbf{G}\mathbf{w}_i\}\mathbf{h}_{\mathrm{r},k},
\end{align}
\end{subequations}
where $\hat{i}$ represents the element in the set $\{1,2\}$ other than $i$, i.e., if $i = 1$ then $\hat{i} = 2$ and if $i = 2$ then $\hat{i} = 1$.

To handle the non-convex fractional terms in (\ref{pr:phi_10}a), we propose to introduce two auxiliary variables $t_1$ and $t_2$ to replace them, as shown below:
\begin{subequations}\label{pr:phi_2}
\allowdisplaybreaks[4]
\begin{align}
\min_{\boldsymbol{\phi}, t_1, t_2} ~~ &t_1 + t_2 - \mathbf{v}^H \mathbf{F}\mathbf{v}\\
\text{s.t.} ~~~~&\boldsymbol{\phi}^H\mathbf{J}\boldsymbol{\phi}\boldsymbol{\phi}^H\boldsymbol{\phi} + \sigma_\mathrm{t}^2 \sigma_\mathrm{z}^2 \alpha_{\mathrm{r,t}}^4(\boldsymbol{\phi}^H\boldsymbol{\phi})^2 \\
&\quad + \boldsymbol{\phi}^H\mathbf{K}\boldsymbol{\phi} \leq P_{\mathrm{RIS}}, \notag\\
& \boldsymbol{\phi}^H\mathbf{C}_k\boldsymbol{\phi} + \Re\{\mathbf{d}_k^H\boldsymbol{\phi}\} + {c}_{\mathrm{\phi},k}\\
 &\quad - (1+{\gamma_k^{-1}})\boldsymbol{\phi}^H\mathbf{b}_{k,k}^{*}\mathbf{b}_{k,k}^{T}\boldsymbol{\phi} \leq 0,~\forall k, \notag\\
&t_i \geq \frac{\bolds{\xi}_i^H\mathbf{v}\mathbf{v}^H \mathbf{\Xi}_i\mathbf{v}}{\boldsymbol{\phi}^H\mathbf{R}_{\hat{i}}\boldsymbol{\phi}},~\forall i, ~\hat{i} \neq i,\\
& a_m \leq a_{\max},~\forall m.
\end{align}
\end{subequations}
Obviously, when $\bolds{\phi}$ is given, the problem (\ref{pr:phi_2}) becomes relevant only for the auxiliary variables $t_1$ and $t_2$. Thus, the optimal solutions $t_1$ and $t_2$ in each iteration can be obtained as
\begin{equation}\label{eq:t}
t_i^{\star} = \frac{\bolds{\xi}_i^H\mathbf{v}\mathbf{v}^H \mathbf{\Xi}_i\mathbf{v}}{\boldsymbol{\phi}^H\mathbf{R}_{\hat{i}}\boldsymbol{\phi}},~\forall i, ~\hat{i} \neq i.\\
\end{equation}

Furthermore, with optimal $t_1$ and $t_2$, the optimization on $\bolds{\phi}$ can be formulated as solving the following problem:
\begin{subequations}\label{pr:phi_3}
\allowdisplaybreaks[4]
\begin{align}
\min_{\boldsymbol{\phi}} ~~ &- \mathbf{v}^H \mathbf{F}\mathbf{v} \label{phi_3a}\\
\text{s.t.} ~~~&\boldsymbol{\phi}^H\mathbf{J}\boldsymbol{\phi}\boldsymbol{\phi}^H\boldsymbol{\phi} + \sigma_\mathrm{t}^2 \sigma_\mathrm{z}^2 \alpha_{\mathrm{r,t}}^4(\boldsymbol{\phi}^H\boldsymbol{\phi})^2 \label{phi_3b}\\
&\quad + \boldsymbol{\phi}^H\mathbf{K}\boldsymbol{\phi} \leq P_{\mathrm{RIS}}, \notag\\
& \boldsymbol{\phi}^H\mathbf{C}_k\boldsymbol{\phi} + \Re\{\mathbf{d}_k^H\boldsymbol{\phi}\} + {c}_{\mathrm{\phi},k} \label{phi_3c}\\
 &\quad - (1+{\gamma_k^{-1}})\boldsymbol{\phi}^H\mathbf{b}_{k,k}^{*}\mathbf{b}_{k,k}^{T}\boldsymbol{\phi} \leq 0,~\forall k,\notag\\
& \bolds{\xi}_i^H\mathbf{v}\mathbf{v}^H \mathbf{\Xi}_i\mathbf{v} - t_i \boldsymbol{\phi}^H\mathbf{R}_{\hat{i}}\boldsymbol{\phi} \leq 0,~\forall i, ~\hat{i} \neq i,\label{phi_3d}\\
& a_m \leq a_{\max},~\forall m.
\end{align}
\end{subequations}
Now, the design problem with regard to active RIS reflection coefficients $\bolds{\phi}$ is relatively more straightforward. Nevertheless, due to the existence of the quartic terms with respect to $\bolds{\phi}$ in (\ref{phi_3a}) and (\ref{phi_3b}) (i.e., $\mathbf{v}^H \mathbf{F}\mathbf{v}$, $\boldsymbol{\phi}^H\mathbf{J}\boldsymbol{\phi}\boldsymbol{\phi}^H\boldsymbol{\phi} $ and $(\boldsymbol{\phi}^H\boldsymbol{\phi})^2 $), the non-convex constraint in (\ref{phi_3c}) and the sextic terms with respect to $\bolds{\phi}$ in (\ref{phi_3d}) (i.e., $\bolds{\xi}_i^H\mathbf{v}\mathbf{v}^H \mathbf{\Xi}_i\mathbf{v}$), problem (\ref{pr:phi_3}) is very tough to deal with. To tackle this difficulty, MM algorithm is utilized to find a series of convex tractable surrogate functions for these non-convex terms via both first-order Taylor expansion and second-order Taylor expansion, as presented in follows.

\textbf{Transformation for Objective (\ref{phi_3a}):}
Concretely, by using the solution $\bolds{\phi}_s$ obtained in the $s$-th iteration and applying first-order Taylor expansion, we can derive an upper-bound for $-\mathbf{v}^H \mathbf{F}\mathbf{v}$ as
\begin{subequations}\label{eq:vFv1}
\begin{align}
  -\mathbf{v}^H \mathbf{F}\mathbf{v} &\leq -\mathbf{v}_{s}^H\mathbf{F}\mathbf{v}_{s}
  - 2\Re\{\mathbf{v}_{s}^H\mathbf{F}(\mathbf{v}-\mathbf{v}_{s})\},\\
  &= \Re\{\mathbf{v}^H \mathbf{f}\} + c_1,\\
  &= \Re\{\bolds{\phi}^H \widetilde{\mathbf{F}}\bolds{\phi}\} + c_1,
\end{align}
\end{subequations}
where we define $\mathbf{f} \eqdef -2\mathbf{F}^H\mathbf{v}_{s}$ and $c_1 \eqdef \mathbf{v}_{s}^H\mathbf{F}\mathbf{v}_{s}$ is a constant independent of $\bolds{\phi}$. Besides, $\widetilde{\mathbf{F}}$ is a reshaped matrix form corresponding to the vector $\mathbf{f}$, that is, $\mathbf{f}=\text{vec}\{\widetilde{\mathbf{F}}\}$. Nevertheless, the real-valued function $\Re\{\bolds{\phi}^H \widetilde{\mathbf{F}}\bolds{\phi}\}$ is still non-convex. Moreover, we suggest re-writing expression $\Re\{\bolds{\phi}^H \widetilde{\mathbf{F}}\bolds{\phi}\}$ in the form of real-valued variables and further find a tractable upper-bound for it via the second-order Taylor expansion.
In particular, with the definitions
\begin{subequations}\label{eq:phi_real}
\begin{align}
  \bar{\boldsymbol{\phi}} &\triangleq \left[\Re\{\boldsymbol{\phi}^{T}\} ~~ \Im\{\boldsymbol{\phi}^{T}\}\right]^{T},\\
  \bar{\mathbf{F}} &\triangleq \begin{bmatrix}
\Re\{\widetilde{\mathbf{F}}\} & -\Im\{\widetilde{\mathbf{F}}\} \\
\Im\{\widetilde{\mathbf{F}}\} & \Re\{\widetilde{\mathbf{F}}\}
\end{bmatrix},
\end{align}
\end{subequations}
we have
\begin{subequations}\label{eq:vFv2}
\allowdisplaybreaks[4]
\begin{align}
  \Re\{\bolds{\phi}^H \widetilde{\mathbf{F}}\bolds{\phi}\} &= \bar{\bolds{\phi}}^T\bar{\mathbf{F}}\bar{\bolds{\phi}},\\
  &\leq \bar{\boldsymbol{\phi}}_s^T \bar{\mathbf{F}} \bar{\boldsymbol{\phi}}_s + \bar{\boldsymbol{\phi}}_s^T (\bar{\mathbf{F}} + \bar{\mathbf{F}}^T)(\bar{\boldsymbol{\phi}} - \bar{\boldsymbol{\phi}}_s)\\
   &\quad + \frac{\widetilde{\lambda}_1}{2}(\bar{\boldsymbol{\phi}} - \bar{\boldsymbol{\phi}}_s)^T(\bar{\boldsymbol{\phi}} - \bar{\boldsymbol{\phi}}_s),\notag\\
  &= \frac{\widetilde{\lambda}_1}{2}\bolds{\phi}^H\bolds{\phi} + \Re\{\bolds{\phi}^H\widetilde{\mathbf{f}}\} -\bar{\bolds{\phi}}_s^T\bar{\mathbf{F}}^T\bar{\bolds{\phi}}_s + \frac{\widetilde{\lambda}_1}{2}\bar{\bolds{\phi}}_s^T\bar{\bolds{\phi}}_s,
\end{align}
\end{subequations}
in which we define $\widetilde{\lambda}_1$ as the maximum eigenvalue of Hessian matrix $(\bar{\mathbf{F}} + \bar{\mathbf{F}}^T)$, $\widetilde{\mathbf{f}} \eqdef {\mathbf{U}} (\bar{\mathbf{F}} + \bar{\mathbf{F}}^T - \widetilde{\lambda}_1\mathbf{I}_{2M})\bar{\bolds{\phi}}_s$ and ${\mathbf{U}} \eqdef \left[\mathbf{I}_{M} ~ \jmath\mathbf{I}_{M}\right]$. By substituting (\ref{eq:vFv2}) into (\ref{eq:vFv1}), a convex surrogate function of $-\mathbf{v}^H \mathbf{F}\mathbf{v}$ can be obtained as
\begin{equation}\label{eq:vFv3}
 -\mathbf{v}^H \mathbf{F}\mathbf{v} \leq \frac{\widetilde{\lambda}_1}{2}\bolds{\phi}^H\bolds{\phi} + \Re\{\bolds{\phi}^H\widetilde{\mathbf{f}}\} + c_2,
\end{equation}
where $c_2 \eqdef -\bar{\bolds{\phi}}_s^T\bar{\mathbf{F}}^T\bar{\bolds{\phi}}_s + \frac{\widetilde{\lambda}_1}{2}\bar{\bolds{\phi}}_s^T\bar{\bolds{\phi}}_s + c_1$.

\textbf{Transformation for Constraint (\ref{phi_3b}):}
Now, the objective (\ref{phi_3a}) is tractable, we then consider handling the constraints (\ref{phi_3b})-(\ref{phi_3d}). Thanks to the amplitude constraint of active RIS, i.e., $a_m \leq a_{\max}$, $\bolds{\phi}^H\bolds{\phi}$ is upper-bounded by $\bolds{\phi}^H\bolds{\phi} \leq Ma_{\max}^2$.
Therefore, the power constraint of the active RIS can be written as
\begin{equation}\label{eq:P_phi_final}
\begin{aligned}
  & \boldsymbol{\phi}^H\mathbf{J}\boldsymbol{\phi}\boldsymbol{\phi}^H\boldsymbol{\phi} + \sigma_\mathrm{t}^2 \sigma_\mathrm{z}^2 \alpha_{\mathrm{r,t}}^4(\boldsymbol{\phi}^H\boldsymbol{\phi})^2 + \boldsymbol{\phi}^H\mathbf{K}\boldsymbol{\phi} \\
    &\quad \leq\boldsymbol{\phi}^H \widetilde{\mathbf{K}} \boldsymbol{\phi} + c_{3} \leq P_\mathrm{RIS},
\end{aligned}
\end{equation}
with $\widetilde{\mathbf{K}} \eqdef \mathbf{K} +Ma_{\max}^2\mathbf{J}$ and $c_3 \eqdef \sigma_\mathrm{t}^2 \sigma_\mathrm{z}^2 \alpha_{\mathrm{r,t}}^4 M^2a_{\max}^4$.

\textbf{Transformation for Constraint (\ref{phi_3c}):} It is obvious that the presence of the concave term $-(1+{\gamma_k^{-1}})\boldsymbol{\phi}^H\mathbf{b}_{k,k}^{*}\mathbf{b}_{k,k}^{T}\boldsymbol{\phi}$ causes the constraint (\ref{phi_3c}) to be non-convex. Particularly, a linear surrogate function for it can be derived as
\begin{equation}\label{eq:bb}
\begin{aligned}
-\boldsymbol{\phi}^H\mathbf{b}_{k,k}^{*}\mathbf{b}_{k,k}^{T}\boldsymbol{\phi} &\leq - \boldsymbol{\phi}_{s}^H\mathbf{b}_{k,k}^{*}\mathbf{b}_{k,k}^{T}\boldsymbol{\phi}_{s}\\
&\quad  - 2\Re\{\boldsymbol{\phi}_{s}^H\mathbf{b}_{k,k}^{*}\mathbf{b}_{k,k}^{T}(\boldsymbol{\phi} -\boldsymbol{\phi}_{s})\}.
\end{aligned}
\end{equation}
On the basis of the result in (\ref{eq:bb}), we can obtain an upper-bound function for $-(1+{\gamma_k^{-1}})\boldsymbol{\phi}^H\mathbf{b}_{k,k}^{*}\mathbf{b}_{k,k}^{T}\boldsymbol{\phi}$ and re-arrange the SINR constraint as
\begin{equation}\label{eq:SINR2}
\boldsymbol{\phi}^H\mathbf{C}_k\boldsymbol{\phi} + \Re\{\widetilde{\mathbf{d}}_k^H\boldsymbol{\phi}\} + \tilde{{c}}_{\mathrm{\phi},k} \leq 0,~\forall k,
\end{equation}
where for brevity we define $\widetilde{\mathbf{d}}_k^H \eqdef {\mathbf{d}}_k^H - 2(1+{\gamma_k^{-1}})\boldsymbol{\phi}_{s}^H\mathbf{b}_{k,k}^{*}\mathbf{b}_{k,k}^{T}$ and $\tilde{{c}}_{\mathrm{\phi},k} \eqdef {{c}}_{\mathrm{\phi},k} + (1+{\gamma_k^{-1}}) \boldsymbol{\phi}_{s}^H\mathbf{b}_{k,k}^{*}\mathbf{b}_{k,k}^{T} \boldsymbol{\phi}_{s}$.

\textbf{Transformation for Constraint (\ref{phi_3d}):} Recalling the definition $\bar{\boldsymbol{\phi}} \triangleq \left[\Re\{\boldsymbol{\phi}^{T}\} ~~ \Im\{\boldsymbol{\phi}^{T}\}\right]^{T}$ and defining other notations for brevity as follows:
\begin{subequations}\label{eq:FP_trans}
\begin{align}
  \bar{\bolds{\xi}}_i &\eqdef \left[\Re\{\boldsymbol{\xi}_i^{T}\} ~~ \Im\{\boldsymbol{\xi}_i^{T}\}\right]^{T},\\
  \bar{\mathbf{v}} &\triangleq \left[\Re\{\mathbf{v}^{T}\} ~~ \Im\{\mathbf{v}^{T}\}\right]^{T},\\
  \bar{\mathbf{\Xi}}_i &\eqdef \begin{bmatrix}
\Re\{\widetilde{{\mathbf{\Xi}}}_i\} & -\Im\{\widetilde{{\mathbf{\Xi}}}_i\} \\
\Im\{\widetilde{{\mathbf{\Xi}}}_i\} & \Re\{\widetilde{{\mathbf{\Xi}}}_i\}
\end{bmatrix},
\end{align}
\end{subequations}
the equivalent real-valued form $y_i(\bar{\mathbf{v}})$ of the first term $\bolds{\xi}_i^H\mathbf{v}\mathbf{v}^H \mathbf{\Xi}_i\mathbf{v}$ in (\ref{phi_3d}) can be expressed as
\begin{equation}\label{eq:y}
  y_i(\bar{\mathbf{v}}) =  \bolds{\xi}_i^H\mathbf{v}\mathbf{v}^H \mathbf{\Xi}_i\mathbf{v} = \bar{\bolds{\xi}}_i^T\bar{\mathbf{v}}\bar{\mathbf{v}}^T \bar{\mathbf{\Xi}}_i\bar{\mathbf{v}},
\end{equation}
whose first-order and second-order derivatives can be calculated as\begin{subequations}\label{eq:y_der}
\begin{align}
\nabla y_i(\bar{\mathbf{v}}) &= \bar{\bolds{\xi}}_i^T\bar{\mathbf{v}} (\bar{\mathbf{\Xi}}_i + \bar{\mathbf{\Xi}}_i^T)\bar{\mathbf{v}} + \bar{\mathbf{v}}^T \bar{\mathbf{\Xi}}_i\bar{\mathbf{v}}\bar{\bolds{\xi}}_i,\\
\nabla^2 y_i(\bar{\mathbf{v}}) &= (\bar{\mathbf{\Xi}}_i + \bar{\mathbf{\Xi}}_i^T)\bar{\mathbf{v}}\bar{\bolds{\xi}}_i^T + (\bar{\bolds{\xi}}_i\bar{\mathbf{v}}^T + \bar{\mathbf{v}}^T \bar{\bolds{\xi}}_i\mathbf{I}_{2M^2}) (\bar{\mathbf{\Xi}}_i + \bar{\mathbf{\Xi}}_i^T).
\end{align}
\end{subequations}
With $\nabla y_i(\bar{\mathbf{v}})$ and $\nabla^2 y_i(\bar{\mathbf{v}})$ shown in (\ref{eq:y_der}), the upper-bounded surrogate function of $y_i(\bar{\mathbf{v}})$ is obtained by
\begin{subequations}\label{eq:constraint_phi_1_MM}
\begin{align}
  y_i(\bar{\mathbf{v}}) & \leq y_i(\bar{\mathbf{v}}_s) + (\nabla y_i(\bar{\mathbf{v}}_s))^T(\bar{\mathbf{v}} - \bar{\mathbf{v}}_s) \\
  &\quad + \frac{\lambda_{\mathrm{y},i}}{2}(\bar{\mathbf{v}} - \bar{\mathbf{v}}_s)^T(\bar{\mathbf{v}} - \bar{\mathbf{v}}_s)\notag\\
   & = \frac{\lambda_{\mathrm{y},i}}{2} \bar{\mathbf{v}}^T\bar{\mathbf{v}} + \bar{\mathbf{v}}^T \bar{\bolds{\ell}}_i + x_i\\
   & = \frac{\lambda_{\mathrm{y},i}}{2} \mathbf{v}^H\mathbf{v} + \Re\{\mathbf{v}^H {\bolds{\ell}}_i \} + x_i\\
   &\leq  \Re\{\bolds{\phi}^H \mathbf{\Omega}_i \bolds{\phi}\} + \frac{\lambda_{\mathrm{y},i}}{2} M^2a_{\max}^4 + x_i\\
   & = \bar{\bolds{\phi}}^T\bar{\mathbf{\Omega}}_i\bar{\bolds{\phi}} + \frac{\lambda_{\mathrm{y},i}}{2} M^2a_{\max}^4 + x_i\\
  &\leq \bar{\boldsymbol{\phi}}_s^T \bar{\mathbf{\Omega}}_i \bar{\boldsymbol{\phi}}_s + \bar{\boldsymbol{\phi}}_s^T (\bar{\mathbf{\Omega}}_i + \bar{\mathbf{\Omega}}_i^T)(\bar{\boldsymbol{\phi}} - \bar{\boldsymbol{\phi}}_s)\\
   &\quad + \frac{\widetilde{\lambda}_{\mathrm{y},i}}{2}(\bar{\boldsymbol{\phi}} - \bar{\boldsymbol{\phi}}_s)^T(\bar{\boldsymbol{\phi}} - \bar{\boldsymbol{\phi}}_s) + \frac{\lambda_{\mathrm{y},i}}{2} M^2a_{\max}^4 + x_i \notag\\
  &= \frac{\widetilde{\lambda}_{\mathrm{y},i}}{2}\bolds{\phi}^H\bolds{\phi} + \Re\{\bolds{\phi}^H\widetilde{\bolds{\ell}}_i\} + \widetilde{x}_i,
\end{align}
\end{subequations}
where $\lambda_{\mathrm{y},i}$ is the maximum eigenvalue of the Hessian matrix $\nabla^2 y_i(\bar{\mathbf{v}}_s)$, $\bar{\bolds{\ell}}_i \eqdef \nabla y_i(\bar{\mathbf{v}}_s) - \lambda_{\mathrm{y},i}\bar{\mathbf{v}}_s$, $x_i \eqdef y_i(\bar{\mathbf{v}}_s) - (\nabla y_i(\bar{\mathbf{v}}_s))^T\bar{\mathbf{v}}_s + \frac{\lambda_{\mathrm{y},i}}{2}\bar{\mathbf{v}}_s^T\bar{\mathbf{v}}_s$, ${\bolds{\ell}}_i \eqdef {\mathbf{U}}_\mathrm{v}\bar{\bolds{\ell}}_i = \text{vec}\{\mathbf{\Omega}_i\}$, ${\mathbf{U}}_\mathrm{v} \eqdef \left[\mathbf{I}_{M^2} ~ \jmath\mathbf{I}_{M^2}\right]$. The inequality (\ref{eq:constraint_phi_1_MM}c)-(\ref{eq:constraint_phi_1_MM}d) holds since $\mathbf{v}^H\mathbf{v} = (\boldsymbol{\phi}^* \otimes  \boldsymbol{\phi})^H(\boldsymbol{\phi}^* \otimes \boldsymbol{\phi})= (\boldsymbol{\phi}^H\boldsymbol{\phi})^2 \leq M^2a_{\max}^4$. In addition, $\bar{\mathbf{\Omega}}_i \eqdef \begin{bmatrix}
\Re\{{{\mathbf{\Omega}}}_i\} & -\Im\{{{\mathbf{\Omega}}}_i\} \\
\Im\{{{\mathbf{\Omega}}}_i\} & \Re\{{{\mathbf{\Omega}}}_i\}
\end{bmatrix}$, $\widetilde{\lambda}_{\mathrm{y},i}$ is the maximum eigenvalue of Hessian matrix $(\bar{\mathbf{\Omega}}_i + \bar{\mathbf{\Omega}}_i^T)$, $\widetilde{\bolds{\ell}}_i \eqdef {\mathbf{U}} (\bar{\mathbf{\Omega}}_i + \bar{\mathbf{\Omega}}_i^T - \widetilde{\lambda}_{\mathrm{y},i}\mathbf{I}_{2M})\bar{\boldsymbol{\phi}}_s$ and $\widetilde{x}_i \eqdef  -\bar{\boldsymbol{\phi}}_s^T\bar{\mathbf{\Omega}}_i^T\bar{\boldsymbol{\phi}}_s + \frac{\widetilde{\lambda}_{\mathrm{y},i}}{2}\bar{\boldsymbol{\phi}}_s^T\bar{\boldsymbol{\phi}}_s + \frac{\lambda_{\mathrm{y},i}}{2} M^2a_{\max}^4 + x_{i}$.

Moreover, a linear surrogate function of $- \boldsymbol{\phi}^H{\mathbf{R}}_{\hat{i}}\boldsymbol{\phi}$ in (\ref{phi_3d}) can be formulated as
\begin{equation}\label{eq:constraint_phi_2_MM}
\begin{aligned}
  - \boldsymbol{\phi}^H{\mathbf{R}}_{\hat{i}}\boldsymbol{\phi} &\leq -\boldsymbol{\phi}_{s}^H{\mathbf{R}}_{\hat{i}}\boldsymbol{\phi}_{s}
  - 2\Re\{\boldsymbol{\phi}_{s}^H{\mathbf{R}}_{\hat{i}}(\boldsymbol{\phi}-\boldsymbol{\phi}_{s})\},\\
  &= \Re\{\boldsymbol{\phi}^H \bolds{\varrho}_i\} + \kappa_i,\\
\end{aligned}
\end{equation}
where we define $\bolds{\varrho}_i \eqdef - 2\mathbf{R}_{\hat{i}}^H \boldsymbol{\phi}_{s}$ and $\kappa_i \eqdef \boldsymbol{\phi}_{s}^H\mathbf{R}_{\hat{i}}\boldsymbol{\phi}_{s}$ for simplicity. To sum up, we obtain the convex surrogate function for constraint (\ref{phi_3d}) as
\begin{equation}\label{eq:constraint}
\bolds{\xi}_i^H\mathbf{v}\mathbf{v}^H \mathbf{\Xi}_i\mathbf{v} - t_i \boldsymbol{\phi}^H\mathbf{R}_{\hat{i}}\boldsymbol{\phi} \leq \frac{\widetilde{\lambda}_{\mathrm{y},i}}{2}\bolds{\phi}^H\bolds{\phi} + \Re\{\bolds{\phi}^H\widetilde{\bolds{\varrho}}_i\} + \widetilde{\kappa}_i,
\end{equation}
with $\widetilde{\bolds{\varrho}}_i \eqdef \widetilde{\bolds{\ell}}_i + t_i\bolds{\varrho}_i$
and $\widetilde{\kappa}_i \eqdef \widetilde{x}_i + t_i\kappa_i$.

Thus, we can formulate the optimization problem with respect to $\bolds{\phi}$ at the $(s+1)$-th iteration as
\begin{subequations}\label{pr:phi_final}
\allowdisplaybreaks[4]
\begin{align}
\min_{\boldsymbol{\phi}} ~~ & \frac{\widetilde{\lambda}_1}{2}\bolds{\phi}^H\bolds{\phi} + \Re\{\bolds{\phi}^H\widetilde{\mathbf{f}}\}\\
\text{s.t.} ~~~
&\boldsymbol{\phi}^H \widetilde{\mathbf{K}} \boldsymbol{\phi} \leq P_{\mathrm{RIS}} - c_{3} ,\\
& \boldsymbol{\phi}^H\mathbf{C}_k\boldsymbol{\phi} + \Re\{\widetilde{\mathbf{d}}_k^H\boldsymbol{\phi}\} + \tilde{{c}}_{\mathrm{\phi},k} \leq 0,~\forall k,\\
&\frac{\widetilde{\lambda}_{\mathrm{y},i}}{2}\bolds{\phi}^H\bolds{\phi} + \Re\{\bolds{\phi}^H\widetilde{\bolds{\varrho}}_i\} + \widetilde{\kappa}_i \leq 0, ~ \forall i,\\
& a_m \leq a_\mathrm{\max},~\forall m.
\end{align}
\end{subequations}
Obviously, it is a simple convex problem that can be readily solved by various convex algorithms/toolboxes.

\subsubsection{Update $\mathbf{\Psi}$}
With optimal $\bolds{\phi}_{s+1}$ in the $(s+1)$-th iteration, we appropriately update the auxiliary variable $\mathbf{\Psi}_{s+1}$ as
\begin{equation}\label{eq:Psi}
  \mathbf{\Psi}_{s+1} = \widetilde{\mathbf{R}}_\mathrm{n} = \sigma_\mathrm{z}^2\mathbf{G}^T\mathbf{\Phi}_{s+1}\mathbf{\Phi}_{s+1}^H\mathbf{G}^* + \sigma_\mathrm{r}^2\mathbf{I}_{N}.
\end{equation}

Finally, by alternatively updating $t_1$, $t_2$, $\bolds{\phi}$ and $\mathbf{\Psi}$, we can solve the active RIS reflection beamforming optimization problem in an iterative manner.

\subsection{Summary, Initialization and Computational Complexity Analysis}
\begin{algorithm}[t]
  \caption{Joint Transmit Precoding and Active RIS Reflection Beamforming Design Algorithm}
  \label{Algorithm 1}
  \begin{algorithmic}[1]
    \REQUIRE $\mathbf{h}_{\mathrm{d},k}$, $\mathbf{h}_{\mathrm{r},k}$, $\mathbf{h}_{\mathrm{r},\mathrm{t}}$, $\mathbf{G}$, $P_\mathrm{BS}$, $P_\mathrm{RIS}$, $\gamma_k$, $a_{\max}$, $\sigma_k$, $\sigma_\mathrm{z}$, $\sigma_\mathrm{r}$, $\sigma_\mathrm{t}$, $N$, $M$, $K$, $L$, $\forall k$.
    \ENSURE  $\mathbf{W}^{\star}$ and $\boldsymbol{\phi}^{\star}$.
    \STATE {Initialize $\boldsymbol{\phi}$.}
    \WHILE{no convergence}
    \STATE {Obtain $\mathbf{W}_i, ~ i = 1, \ldots, K$ and $\mathbf{R}_{\mathrm{w}}$ by solving (\ref{pr:sub_problem_w4});}
    \STATE {Construct $\mathbf{w}_i ,~ i = 1, \ldots, K$ by (\ref{eq:W_final});}
    \STATE {Construct $\mathbf{w}_i,~ i = K+1, \ldots, K+N$ by Cholesky decomposition;}
    \STATE {Combine $\mathbf{W} = [\mathbf{w}_1, \ldots, \mathbf{w}_{K+N}]$.}
        \WHILE{no convergence}
    \STATE {Calculate $t_1$ and $t_2$ by (\ref{eq:t});}
    \STATE {Update $\boldsymbol{\phi}$ by solving (\ref{pr:phi_final});}
    \STATE {Update $\mathbf{\Psi}$ by (\ref{eq:Psi}).}
        \ENDWHILE
    \ENDWHILE
    \STATE {Return $\mathbf{W}^{\star} = \mathbf{W}$ and $\boldsymbol{\phi}^{\star} = \boldsymbol{\phi}$}.
  \end{algorithmic}
\end{algorithm}

According to the above derivations, we summarize the proposed joint transmit precoding and active RIS reflection beamforming design for communication QoS-constrained radar CRB minimization problem in Algorithm \ref{Algorithm 1}. With a suitable initialization, we can iteratively update each variable until convergence.

In general, the performance and convergence speed of the algorithm based on alternating optimization can be influenced by the initialization. It is essential to select a suitable initial point for the optimization problem. Intuitively, the active RIS is employed to improve the wireless propagation environment between the BS and the target/users. Therefore, channel power gains can be regarded as an appropriate performance metric to initialize $\bolds{\phi}$. Specifically, we assume that all the active elements can reach their amplitude maximum $a_{\max}$ and optimize their initialized phase-shifts $\boldsymbol{\psi} \eqdef [\psi_1,\ldots, \psi_M]^T \in \mathbb{C}^M$ as follows:

\begin{equation}\label{pr:initial_phi}
\begin{aligned}
\max_{{\bolds{\psi}}} &~~\|\mathbf{h}_{\mathrm{r,t}}^T\text{diag}\{\bolds{\psi}\}\mathbf{G}\|^2_2 + \sum\nolimits_{k = 1}^{K}\|\mathbf{h}_{\mathrm{d},k}^T + \mathbf{h}_{\mathrm{r},k}^T\text{diag}\{\bolds{\psi}\}\mathbf{G}\|^2_2\\
\text{s.t.} ~&~~|{\psi}_m| = 1, ~\forall m.
\end{aligned}
\end{equation}
Moreover, with the following definitions
\begin{subequations}\label{eq:trans_initial0}
\allowdisplaybreaks[4]
\begin{align}
  \mathbf{M} &\eqdef \text{diag}\{\mathbf{h}_{\mathrm{r,t}}^H\}\mathbf{G}^*\mathbf{G}^T\text{diag}\{\mathbf{h}_{\mathrm{r,t}}\}\\
    &\quad+ \sum\nolimits_{k = 1}^{K} \text{diag}\{\mathbf{h}_{\mathrm{r},k}^H\}\mathbf{G}^*\mathbf{G}^T\text{diag}\{\mathbf{h}_{\mathrm{r},k}\},\notag\\
  \mathbf{m} &\eqdef 2\sum\nolimits_{k = 1}^{K}\text{diag}\{\mathbf{h}_{\mathrm{r},k}^H\}\mathbf{G}^*\mathbf{h}_{\mathrm{d},k},
\end{align}
\end{subequations}
problem (\ref{pr:initial_phi}) is re-formulated as
\begin{equation}\label{pr:initial_phi1}
\begin{aligned}
\min_{{\bolds{\psi}}} &~~-\bolds{\psi}^H\mathbf{M}\bolds{\psi} - \Re\{\bolds{\psi}^H\mathbf{m}\}\\
\text{s.t.} ~&~~|{\psi}_m| = 1, ~\forall m,
\end{aligned}
\end{equation}
which can be efficiently solved by the Riemannian conjugate gradient (RCG) algorithm \cite{LiuR5}. Finally, after having the optimal ${\bolds{\psi}}$ in (\ref{pr:initial_phi1}), the initial $\bolds{\phi}$ can be obtained as $\bolds{\phi} = a_{\max}\bolds{\psi}$, where $a_{\max}$ is the maximum amplitude of the reflection coefficient.

The computational complexity of the proposed CRB optimization algorithm is analyzed as follows, where it is assumed that the popular interior point method is used for solving convex optimization problems in this paper. First of all, initializing active RIS reflection vector $\bolds{\phi}$ requires about $\mathcal{O}(M^{1.5})$ operations. Moreover, obtaining the optimal solutions $\mathbf{W}_i, ~ i = 1,\ldots, K$ and $\mathbf{R}_{\mathrm{w}}$ requires approximately $\mathcal{O}(N^{6.5}K^{6.5})$ operations. The construction of $\mathbf{w}_i, ~ i = 1,\ldots, K+N$ has a computational complexity of $\mathcal{O}((K+N)N^{3})$. The complexity of calculating $t_1$ and $t_2$ is at the order of $\mathcal{O}(M^{2})$. The computational complexity of solving the sub-problem with respect to $\bolds{\phi}$ has the order of $\mathcal{O}(M^{4.5})$. $\mathcal{O}(NM^{2}+ MN^{2})$ operations are demanded to update $\mathbf{\Psi}$. Therefore, the overall computational complexity of Algorithm \ref{Algorithm 1} can be approximated as $\mathcal{O}(N^{6.5}K^{6.5} + M^{4.5} + NM^{2}+ MN^{2})$.

\section{Simulation Results}
\label{4}
\begin{figure}[t]
\centering
  \includegraphics[width = 3.0 in]{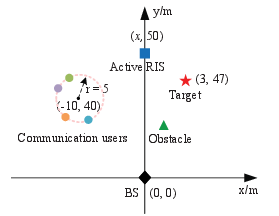}
  \caption{An illustration of the position of BS, active RIS, communication users, and target.}
  \label{fig:position}
  \vspace{-0.2cm}
\end{figure}

In this section, we provide extensive simulation results to verify the advantages of the proposed active RIS-empowered ISAC design scheme. As illustrated in Fig. \ref{fig:position}, we assume that a dual-functional BS with $N = 16$ transmit/receive antennas is at the origin of coordinates and performs communication and sensing functions at the same time. The communication users are located in a circle with the center at (-10m, 40m) and a radius of 5m. Meanwhile, a potential detection target blocked by obstacles is situated in (3m, 47m). The ISAC system is assisted by an active RIS located at ($x$ = 0m, 50m)\footnote{It is worth noting that both communication users and sensing target are located in the far-field of the BS and RIS.}. In light of the trade-off between performance and cost/energy efficiency considerations, we have set the number of active RIS reflection elements to 8 in this work. It is assumed that the BS-RIS channel and the BS/RIS-user channels follow the Rician fading model and Rayleigh fading model, respectively. The RIS-target channel is assumed to be the LoS model, as defined before. In specific, the DoA of the target with respect to active RIS is set as $\theta = \frac{\pi}{4}$. The typical path-loss model $PL(d) = C_0(d_0/d)^{\iota}$ is adopted in this paper. The path-loss exponents for the above channels are set to 2.2, 3.5, 2.3, and 2.2, respectively.
Besides, it is assumed that all communication users have the same QoS requirements for the sake of simplicity, i.e., $\gamma_k = \gamma, ~\forall k$. Finally, the noise powers are set to $\sigma_k^2 = \sigma_\mathrm{z}^2 = \sigma_\mathrm{r}^2 = -80$dBm,$~\forall k$, the RCS is set to $\sigma_\mathrm{t}^2 = 1$, and the number of samples is set to $L = 1024$.

\begin{figure}[t]
\centering
     \begin{minipage}[t]{0.45\linewidth}
        \centering
        \includegraphics[width=\textwidth]{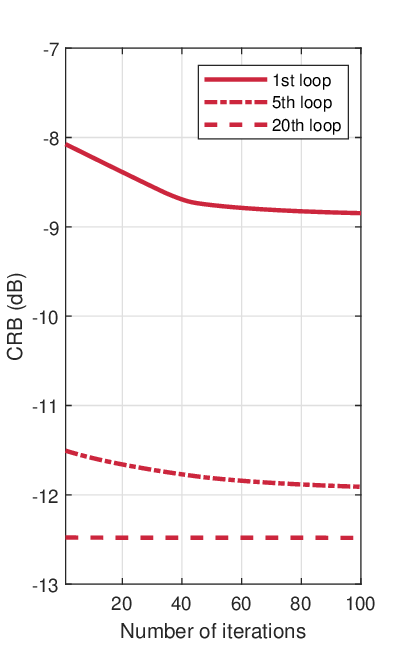}
        \centerline{(a) Inner loop.}
    \end{minipage}%
    \begin{minipage}[t]{0.45\linewidth}
        \centering
        \includegraphics[width=\textwidth]{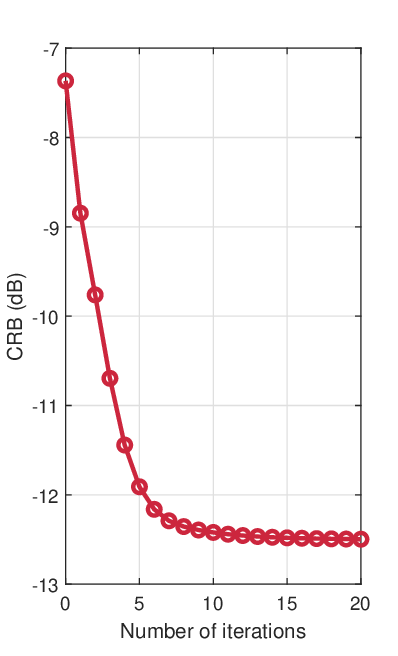}
        \centerline{(b) Outer loop.}
    \end{minipage}
    \caption{Convergence of Algorithm \ref{Algorithm 1}.}
  \label{fig:convergence}
  \vspace{-0.2cm}
\end{figure}

The convergence behavior of Algorithm \ref{Algorithm 1} is illustrated in Fig. \ref{fig:convergence}, where we set $P_{\mathrm{BS}}$ = 27dBm, $P_{\mathrm{RIS}}$ = 10dBm, $a_{\max}$ = 8, and $\gamma$ = 16dB. Specifically, Fig. \ref{fig:convergence}(a) and Fig. \ref{fig:convergence}(b) show the convergence of the inner active RIS reflection beamforming design algorithm and the outer whole algorithm, respectively.
Particularly, the curves in Fig. \ref{fig:convergence}(a) represent the inner convergence of the 1st, 5th, and 20th outer iterations, respectively, i.e., corresponding to the 1st, 5th, and 20th points in Fig. \ref{fig:convergence}(b).
It is observed that the proposed algorithm converges within a finite number of iterations and exhibits excellent convergence performance.

\begin{figure}[t]
	\centering
	\includegraphics[width = 3.6 in]{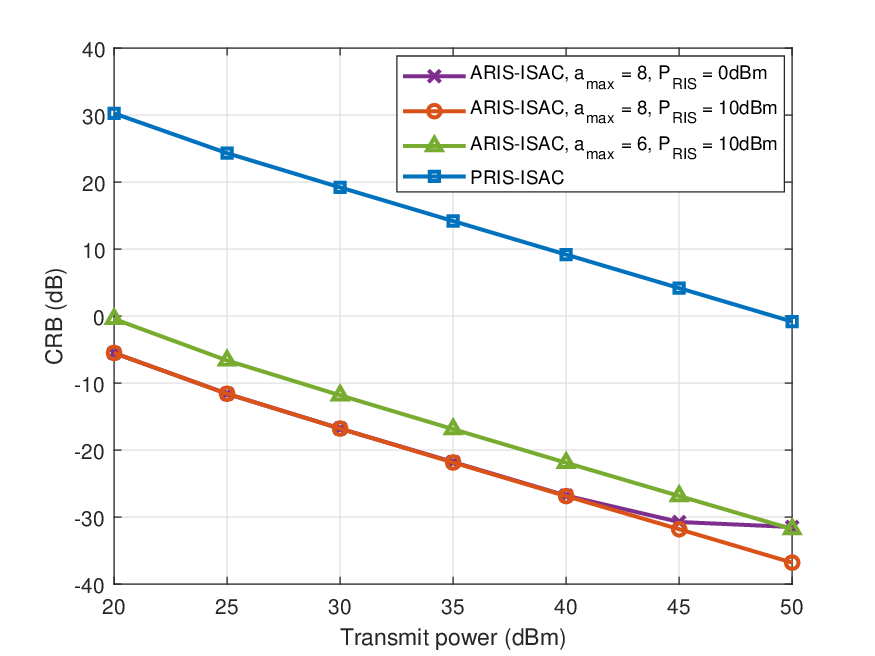}
	\caption{CRB versus the transmit power $P_\mathrm{BS}$ ($\gamma$ = 16dB).}
	\label{fig:Pb}
	\vspace{-0.2cm}
\end{figure}

We then present the CRB for target DoA estimation versus transmit power $P_{\mathrm{BS}}$ in Fig. \ref{fig:Pb}. In order to demonstrate the advantages of the proposed active RIS-empowered ISAC scheme (denoted as \textbf{``ARIS-ISAC"}), the passive RIS-aided ISAC scheme is included as a benchmark (denoted as \textbf{``PRIS-ISAC"}). To achieve a fair comparison, we guarantee that the total power budgets are the same for the active RIS and passive RIS systems, i.e., $P_{\mathrm{BS}}^{\mathrm{p}} = P_{\mathrm{BS}} + P_{\mathrm{RIS}}$ ($P_{\mathrm{RIS}}$ = 10dBm) is set as transmit power for passive RIS scheme. It can be easily observed that the proposed active RIS-empowered ISAC scheme is consistently superior to the passive one for all transmit powers, and achieves up to 36dB CRB performance improvement at the case of $a_{\max}$ = 8, $P_{\mathrm{RIS}}$ = 10dBm, which validates the significant benefits of deploying active RIS in ISAC systems compared to passive RIS. In fact, the drawback of employing passive RIS is more pronounced in sensing a target located in a blind area of BS, since the echo signals suffer from severe fading of BS-RIS-target channel twice (forward and backward). Weak echo signals prevent us from extracting useful information on target detection/parameter estimation. Fig. \ref{fig:Pb} indicates that by favorably amplifying the echo signals twice, active RIS can successfully overcome this multiplicative fading effect. Furthermore, for given maximum amplitude factor $a_{\max} = 8$, we can notice that the active RIS schemes with different $P_{\mathrm{RIS}}$ budgets have almost the same CRB when $P_{\mathrm{BS}}$ is small. It implies that for the weak transmit power, the active RIS amplitude constraint is dominant while the active RIS power constraint is inactive. As $P_{\mathrm{BS}}$ increases, curves with different $P_{\mathrm{RIS}}$ settings are distinguished from the other and the CRB performance at the case of $P_{\mathrm{RIS}}$ = 0dBm tends to be saturated. This phenomenon reveals that in the strong transmit power scenario, the active RIS power constraint further limits the radar performance. Besides, the case of $a_{\max} = 8$, $P_{\mathrm{RIS}}$ = 10dBm is always superior to the case of $a_{\max} = 6$, $P_{\mathrm{RIS}}$ = 10dBm, which means that a wider range of amplitude variation can bring a higher DoF and lead to better performance when the active RIS power budget is sufficient.

\begin{figure}[t]
\centering
  \includegraphics[width = 3.6 in]{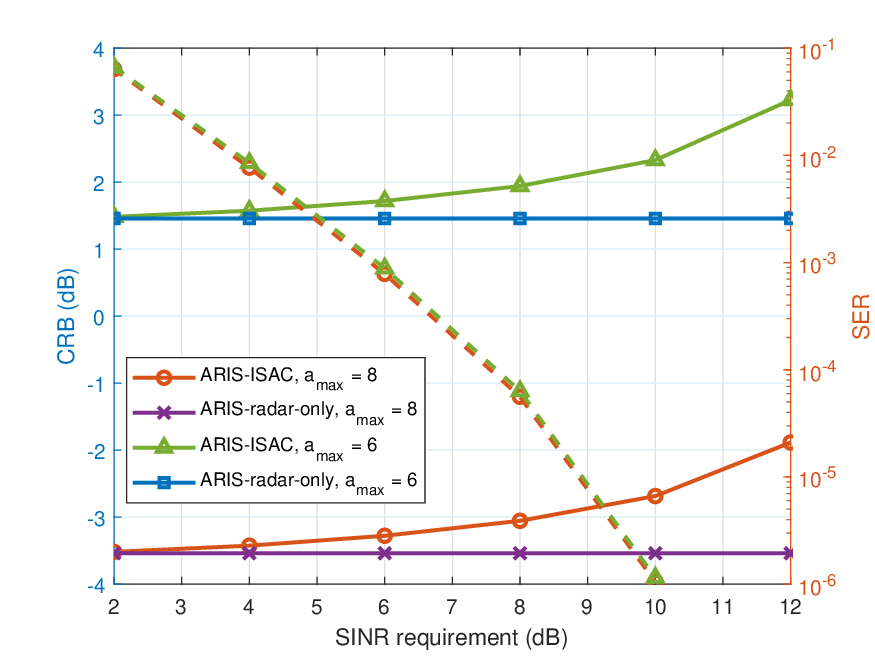}
  \caption{CRB/SER versus the SINR requirement $\gamma$ ($P_{\mathrm{BS}}$ = 16dBm, $P_\mathrm{RIS}$ = 10dBm, solid line: CRB, dashed line: SER).}
  \label{fig:gamma}
  \vspace{-0.2cm}
\end{figure}

\begin{figure}[t]
\centering
  \includegraphics[width = 3.6 in]{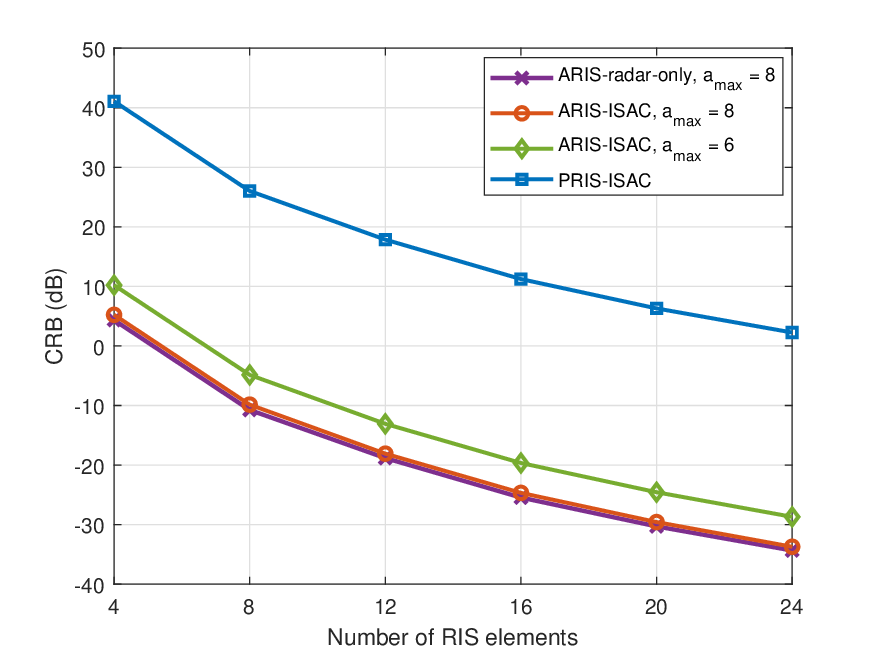}
  \caption{CRB versus the number of RIS elements $M$ ($P_{\mathrm{BS}}$ = 23dBm, $P_\mathrm{RIS}$ = 10dBm, $\gamma$ = 16dB).}
  \label{fig:M}
  \vspace{-0.2cm}
\end{figure}

\begin{figure}[t]
\centering
  \includegraphics[width = 3.6 in]{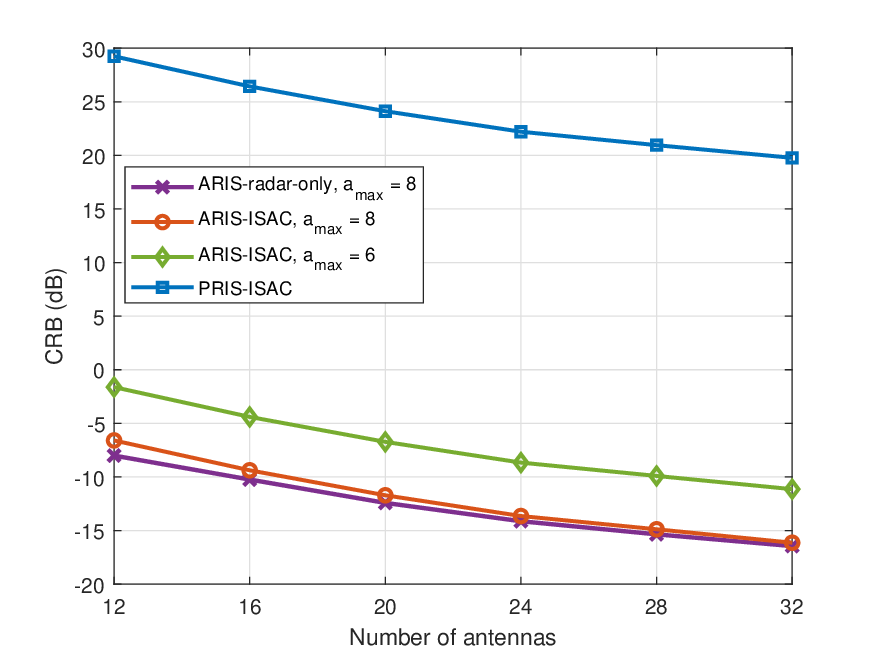}
  \caption{CRB versus the number of antennas $N$ ($P_{\mathrm{BS}}$ = 23dBm, $P_\mathrm{RIS}$ = 10dBm, $\gamma$ = 16dB).}
  \label{fig:N}
  \vspace{-0.2cm}
\end{figure}

The CRB/symbol error rate (SER) performance versus the communication users' SINR requirement $\gamma$ is studied in Fig. \ref{fig:gamma}, in which the solid lines represent the CRB performance for sensing function, while the dashed lines represent the SER performance for communication function. Firstly, not surprisingly, we can observe that the SER of the active RIS-assisted ISAC system gradually decreases as it has higher communication SINR requirements, which means that the communication QoS can be satisfied.
In addition, the active RIS-aided radar-only system is considered as a baseline (denoted as \textbf{``ARIS-radar-only"}). In comparison to the radar-only system, the active RIS-assisted ISAC system incurs a certain CRB performance loss as it has higher communication requirements. This loss is insignificant when $\gamma$ is small, since the beamforming solutions obtained by minimizing the radar CRB can satisfy the communication SINR. With the growth of $\gamma$, the performance gap between the ISAC system and the radar-only system becomes more evident. This is because more resources are skewed toward communication function, resulting in a rise of the CRB for target estimation, which proves the trade-off between multi-user communications and radar sensing on ISAC systems.

The CRB versus the number of RIS reflection elements is demonstrated in Fig. \ref{fig:M}.
We can observe that the proposed CRB minimization for the active RIS-empowered ISAC system dramatically outperforms the passive RIS-assisted system and has quite close performance to the active RIS-empowered radar-only system. As expected, the CRB of all scenarios decreases with the increase of $M$ owing to higher exploitable spatial DoFs.

\begin{figure}[t]
\centering
  \includegraphics[width = 3.6 in]{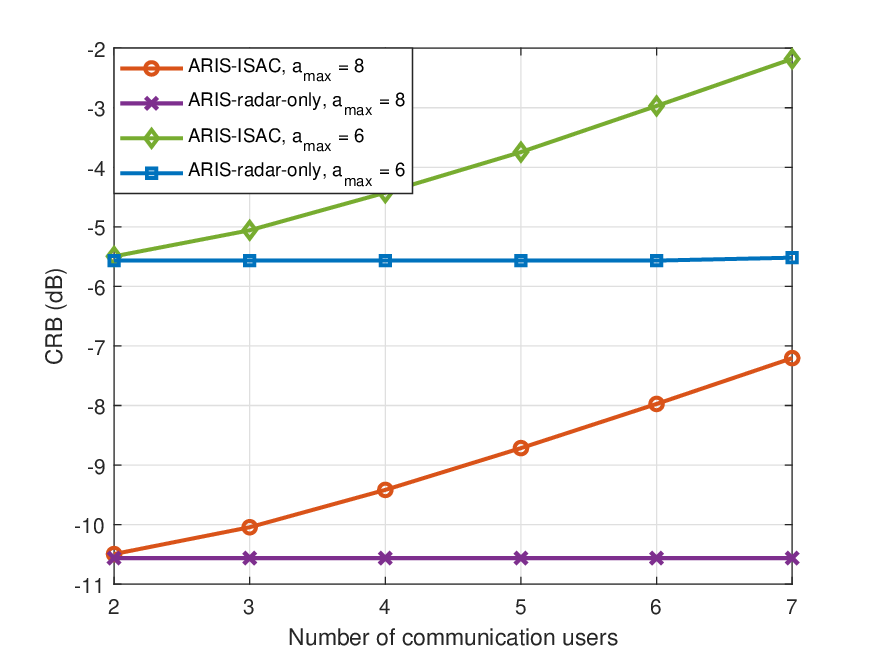}
  \caption{CRB versus the number of communication users $K$ ($P_{\mathrm{BS}}$ = 23dBm, $P_\mathrm{RIS}$ = 10dBm, $\gamma$ = 16dB).}
  \label{fig:K}
  \vspace{-0.2cm}
\end{figure}

\begin{figure}[t]
\centering
  \includegraphics[width = 3.6 in]{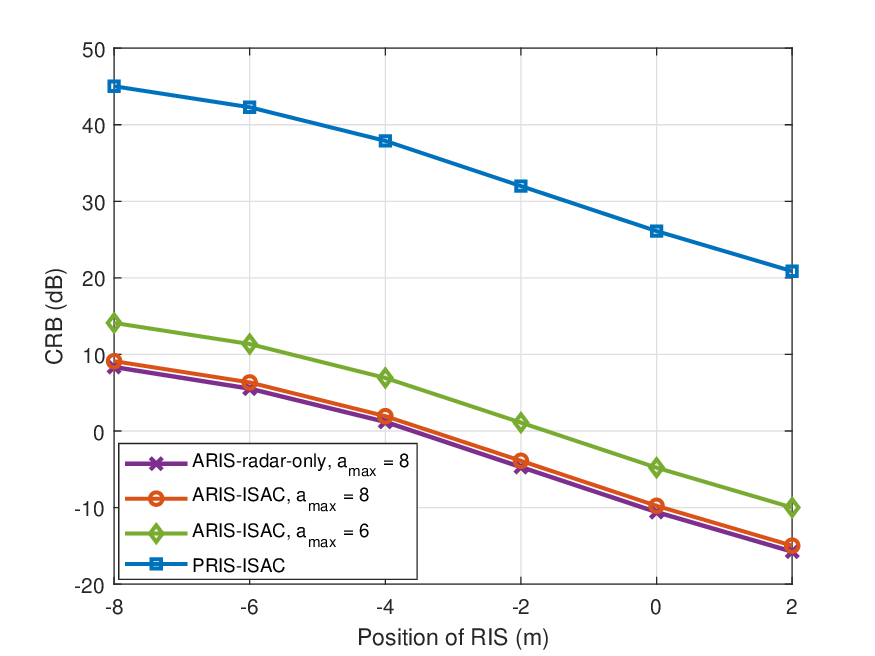}
  \caption{CRB versus the position of RIS $x$ ($P_{\mathrm{BS}}$ = 23dBm, $P_\mathrm{RIS}$ = 10dBm, $\gamma$ = 16dB).}
  \label{fig:dis2}
  \vspace{-0.2cm}
\end{figure}

Next, we illustrate the CRB performance versus the number of antennas $N_\mathrm{t} = N_\mathrm{r} = N$ in Fig. \ref{fig:N}. Similar conclusions can be drawn from Fig. \ref{fig:N} that the active RIS solution performs better than the passive RIS solution and the active RIS-empowered ISAC system behaves very similarly to the active RIS-empowered radar-only system. In addition, improved performance can be obtained by adding antennas owing to more spatial diversity and larger beamforming gains. Furthermore, it is worth noting that as $N$ grows, the performance of the ISAC system and the radar-only system gradually approaches the same due to the limitation of the active RIS power budget.

\begin{figure}[t]
\centering
  \includegraphics[width = 3.64 in]{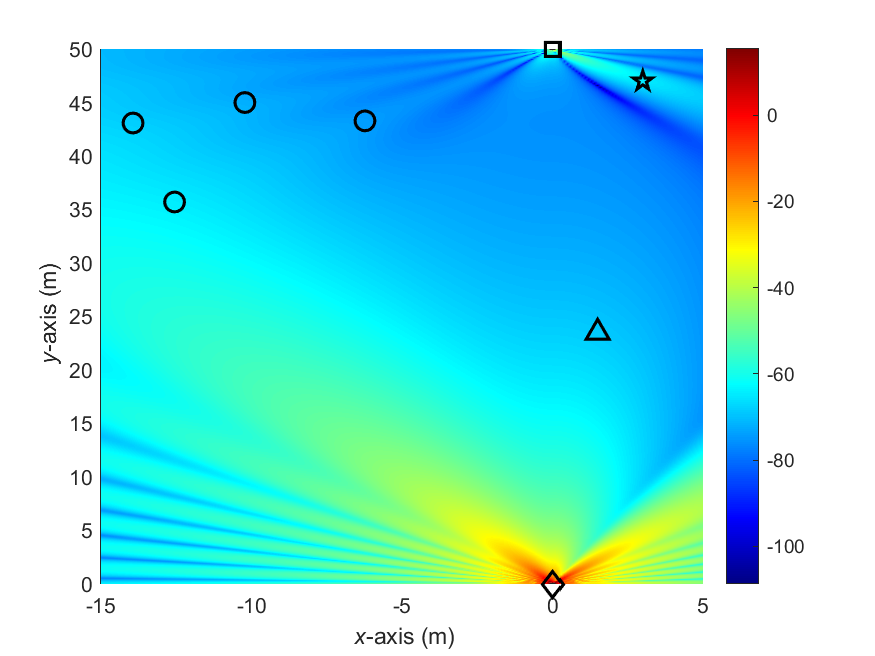}
  \caption{Beampattern of the active RIS-assisted system (BS: diamond; RIS: square; target: star; users: circles, obstacle: triangle).}
  \label{fig:beam}
  \vspace{-0.2cm}
\end{figure}

In Fig. \ref{fig:K}, we display the CRB for target DoA estimation as a function of the number of communication users $K$. As we can predict, with growing $K$, more resources in the ISAC system are allocated to the communication function to ensure the QoS for communication users, which causes a deterioration of the radar sensing function, i.e., an increase of CRB. We also plot the CRB versus the position of RIS in Fig. \ref{fig:dis2}. Obviously, the closer the distance between the RIS and the target, the better the sensing performance of the radar can be achieved.

In Fig. \ref{fig:beam}, we present the illustration of the beampattern of the active RIS-assisted ISAC system in the case of LoS channels. It can be observed that the BS transmit beams are strongly directed toward the active RIS and the communication users.
More importantly, the reflection beams of the active RIS are also pointed toward the target and the users to achieve satisfactory communication and sensing performance.



\section{Conclusions}
In this paper, we investigated the joint transmit precoding and active RIS reflection beamforming design for active RIS-empowered ISAC systems. In addition to the SINR performance metric for multi-user communications, we derived the CRB performance metric for evaluating the target DoA estimation to enhance the sensing performance of ISAC systems. Then, we formulated the CRB minimization problem subject to the users' SINR requirements, the BS power budget, the active RIS power budget, and the amplitude constraint of the active RIS reflection coefficients. An effective solution based on alternating optimization, SDR, and MM was exploited to address this extremely challenging problem. Various simulation results verified the effectiveness of the proposed algorithm, illustrated the remarkable performance enhancements from active RIS, and demonstrated the trade-off between multi-user communications and radar sensing. Compared to passive RIS-assisted ISAC system, active RIS can provide more than 30dB reduction of CRB for the single-target DoA estimation. For the case of sensing multiple targets in the active RIS-empowered ISAC system, the derivation of CRB for estimating multiple DoAs and the development of the associated beamforming design algorithms are substantially more complicated, which deserve further investigation in future studies.

\begin{appendices}
\section{}
\label{B}
In order to re-arrange $g(\bolds{\phi})$ as an explicit expression over $\bolds{\phi}$, we first expand and re-write each of the components, which is presented as follows
\begin{subequations}\label{eq:gphis}
\allowdisplaybreaks[4]
\begin{align}
& g_1(\bolds{\phi}) = \text{Tr}\{\dot{\mathbf{Q}}\mathbf{W}\mathbf{W}^H\dot{\mathbf{Q}}^H\mathbf{\Psi}^{-1}\}\\
& = \text{Tr}\{c_0 \mathbf{G}^T {\mathbf{A}}(\mathbf{L}\bolds{\phi}\bolds{\phi}^T + \bolds{\phi}\bolds{\phi}^T\mathbf{L}){\mathbf{A}}\mathbf{G}\mathbf{W}\mathbf{W}^H \\
&\quad \times c_0^* \mathbf{G}^H \mathbf{A}^H(\mathbf{L}\bolds{\phi}^*\bolds{\phi}^H + \bolds{\phi}^*\bolds{\phi}^H\mathbf{L})\mathbf{A}^H\mathbf{G}^* \mathbf{\Psi}^{-1}\} \notag\\
&= |c_0|^2 (\bolds{\phi}^H\mathbf{L}\mathbf{R}_1\bolds{\phi}\bolds{\phi}^H\mathbf{R}_2\mathbf{L}\bolds{\phi} + \bolds{\phi}^H\mathbf{R}_1\bolds{\phi}\bolds{\phi}^H\mathbf{L}\mathbf{R}_2\mathbf{L}\bolds{\phi}
\\
&\quad+\bolds{\phi}^H\mathbf{L}\mathbf{R}_1\mathbf{L}\bolds{\phi}\bolds{\phi}^H\mathbf{R}_2\bolds{\phi}
+\bolds{\phi}^H\mathbf{R}_1\mathbf{L}\bolds{\phi}\bolds{\phi}^H\mathbf{L}\mathbf{R}_2\bolds{\phi}),\notag\\
& g_2(\bolds{\phi}) = |\text{Tr}\{{\mathbf{Q}}\mathbf{W}\mathbf{W}^H\dot{\mathbf{Q}}^H\mathbf{\Psi}^{-1}\}|^2\\
&=|\text{Tr}\{ \alpha_\mathrm{r,t}^2 \mathbf{G}^T\mathbf{A}\bolds{\phi}\bolds{\phi}^T\mathbf{A}\mathbf{G}\mathbf{W}\mathbf{W}^H \\ &\quad \times c_0^* \mathbf{G}^H \mathbf{A}^H(\mathbf{L}\bolds{\phi}^*\bolds{\phi}^H + \bolds{\phi}^*\bolds{\phi}^H\mathbf{L})\mathbf{A}^H\mathbf{G}^* \mathbf{\Psi}^{-1}\}|^2 \notag\\
& = \alpha_\mathrm{r,t}^4 |c_0|^2 (|\bolds{\phi}^H\mathbf{L}\mathbf{R}_1\bolds{\phi}\bolds{\phi}^H\mathbf{R}_2\bolds{\phi}|^2 \\
&\quad+ |\bolds{\phi}^H\mathbf{R}_1\bolds{\phi}\bolds{\phi}^H\mathbf{L}\mathbf{R}_2\bolds{\phi}|^2
\notag \\
&\quad + \bolds{\phi}^H\mathbf{L}\mathbf{R}_1\bolds{\phi}\bolds{\phi}^H\mathbf{R}_1\bolds{\phi}\bolds{\phi}^H\mathbf{R}_2\bolds{\phi}\bolds{\phi}^H\mathbf{R}_2\mathbf{L}\bolds{\phi}
\notag \\ &\quad + \bolds{\phi}^H\mathbf{R}_1\mathbf{L}\bolds{\phi}\bolds{\phi}^H\mathbf{R}_1\bolds{\phi}\bolds{\phi}^H\mathbf{R}_2\bolds{\phi}\bolds{\phi}^H\mathbf{L}\mathbf{R}_2\bolds{\phi}
),\notag\\
& g_3(\bolds{\phi}) = \text{Tr}\{{\mathbf{Q}}\mathbf{W}\mathbf{W}^H{\mathbf{Q}}^H\mathbf{\Psi}^{-1}\}\\
&= \text{Tr}\{\alpha_\mathrm{r,t}^2\mathbf{G}^T\mathbf{A}\bolds{\phi}\bolds{\phi}^T\mathbf{A}\mathbf{G} \mathbf{W}\mathbf{W}^H 	\\	
&\quad \times \alpha_\mathrm{r,t}^2 \mathbf{G}^H\mathbf{A}^H\bolds{\phi}^*\bolds{\phi}^H\mathbf{A}^H\mathbf{G}^*\mathbf{\Psi}^{-1}\} \notag\\
& = \alpha_\mathrm{r,t}^4\bolds{\phi}^H\mathbf{R}_1\bolds{\phi}\bolds{\phi}^H\mathbf{R}_2\bolds{\phi},
\end{align}
\end{subequations}
where we define the Hermitian matrices $\mathbf{R}_1$ and $\mathbf{R}_2$ as
\begin{subequations}\label{eq:phi_0}
\begin{align}
\mathbf{R}_1 &\eqdef \mathbf{A}^H\mathbf{G}^*\mathbf{W}^*\mathbf{W}^T\mathbf{G}^T\mathbf{A},\\
\mathbf{R}_2 &\eqdef \mathbf{A}^H\mathbf{G}^*\mathbf{\Psi}^{-1}\mathbf{G}^T\mathbf{A}.
\end{align}
\end{subequations}
Plugging the results $g_1(\bolds{\phi})$, $g_2(\bolds{\phi})$ and $g_3(\bolds{\phi})$ in (\ref{eq:gphis}), the objective function $g(\bolds{\phi})$ can be re-formulated in an explicit form of the variable $\bolds{\phi}$ as
\begin{subequations}\label{eq:gphi2}
\allowdisplaybreaks[4]
\begin{align}
g(\bolds{\phi})
&= g_1(\bolds{\phi}) - \frac{g_2(\bolds{\phi})}{g_3(\bolds{\phi})}\\
&= |c_0|^2 (\bolds{\phi}^H\mathbf{R}_1\bolds{\phi}\bolds{\phi}^H\mathbf{L}\mathbf{R}_2\mathbf{L}\bolds{\phi}
+\bolds{\phi}^H\mathbf{R}_2\bolds{\phi}\bolds{\phi}^H\mathbf{L}\mathbf{R}_1\mathbf{L}\bolds{\phi} \\
 & \quad - \frac{\bolds{\phi}^H\mathbf{R}_2\bolds{\phi}|\bolds{\phi}^H\mathbf{L}\mathbf{R}_1\bolds{\phi}|^2}{\bolds{\phi}^H\mathbf{R}_1\bolds{\phi}} - \frac{\bolds{\phi}^H\mathbf{R}_1\bolds{\phi}|\bolds{\phi}^H\mathbf{L}\mathbf{R}_2\bolds{\phi}|^2}{\bolds{\phi}^H\mathbf{R}_2\bolds{\phi}} )\notag.
\end{align}
\end{subequations}
Furthermore, by utilizing the properties $\text{Tr}\{\mathbf{A}\mathbf{B}\} = \text{vec}^H\{\mathbf{B}^H\}\text{vec}\{\mathbf{A}\}$ and $\text{Tr}\{\mathbf{A}\mathbf{B}\mathbf{C}\mathbf{D}\} = \text{vec}^H\{\mathbf{D}^H\}(\mathbf{C}^T \otimes \mathbf{A})\text{vec}\{\mathbf{B}\}$, we have
\begin{subequations}\label{eq:AC1}
\allowdisplaybreaks[4]
\begin{align}
 \boldsymbol{\phi}^H\mathbf{R}_i\boldsymbol{\phi} &= \text{Tr}\{\boldsymbol{\phi}\boldsymbol{\phi}^H\mathbf{R}_i\}\\
  &= \text{vec}^H\{\mathbf{R}_i^H\}\text{vec}\{\boldsymbol{\phi}\boldsymbol{\phi}^H\}\notag\\
  &= \bolds{\xi}_i^H\mathbf{v},\notag\\
 |\boldsymbol{\phi}^H\mathbf{L}\mathbf{R}_i\boldsymbol{\phi}|^2
&= \text{Tr}\{\mathbf{L}\mathbf{R}_i\boldsymbol{\phi}\boldsymbol{\phi}^H \mathbf{R}_i\mathbf{L}\boldsymbol{\phi}\}\\
&= \text{vec}^H\{\boldsymbol{\phi}\boldsymbol{\phi}^H\}(\mathbf{L}\mathbf{R}_{{i}}^T \otimes \mathbf{L}\mathbf{R}_i)\text{vec}\{\boldsymbol{\phi}\boldsymbol{\phi}^H\} \notag \\
&= \mathbf{v}^H\mathbf{\Xi}_{\hat{i}}\mathbf{v},\notag\\
  \boldsymbol{\phi}^H\mathbf{R}_i\boldsymbol{\phi}\boldsymbol{\phi}^H\mathbf{L}\mathbf{R}_{\hat{i}}\mathbf{L}\boldsymbol{\phi}&= \text{Tr}\{\mathbf{R}_i\boldsymbol{\phi}\boldsymbol{\phi}^H\mathbf{L}\mathbf{R}_{\hat{i}}\mathbf{L}\boldsymbol{\phi}\boldsymbol{\phi}^H\}\\
  &= \text{vec}^H\{\boldsymbol{\phi}\boldsymbol{\phi}^H\}(\mathbf{L}\mathbf{R}_{\hat{i}}^T\mathbf{L} \otimes \mathbf{R}_i)\text{vec}\{\boldsymbol{\phi}\boldsymbol{\phi}^H\} \notag \\
  &  = \mathbf{v}^H \mathbf{F}_i\mathbf{v}\notag,
\end{align}
\end{subequations}
where for conciseness we define
\begin{subequations}\label{eq:phi}
\allowdisplaybreaks[4]
\begin{align}
\mathbf{v} &\eqdef \text{vec}\{\boldsymbol{\phi}\boldsymbol{\phi}^H\} = \boldsymbol{\phi}^* \otimes \boldsymbol{\phi},\\
\boldsymbol{\xi}_i &\eqdef \text{vec}\{\mathbf{R}_i^H\}, ~ i = 1,2,\\
\mathbf{\Xi}_i &\eqdef \mathbf{L}\mathbf{R}_{{\hat{i}}}^T \otimes \mathbf{L}\mathbf{R}_{{\hat{i}}},~\forall i, ~\hat{i} \neq i,\\
\mathbf{F}_i &\eqdef \mathbf{L}\mathbf{R}_{\hat{i}}^T\mathbf{L} \otimes \mathbf{R}_i,~\forall i, ~\hat{i} \neq i,
\end{align}
\end{subequations}
and $\hat{i}$ represents the element in the set $\{1,2\}$ other than $i$, i.e., if $i = 1$ then $\hat{i} = 2$ and if $i = 2$ then $\hat{i} = 1$. Then, submitting the transformations in (\ref{eq:AC1}) into (\ref{eq:gphi2}) and defining $\mathbf{F} \eqdef \mathbf{F}_1 + \mathbf{F}_2$, the objective of optimization problem (\ref{pr:objective_phi_1}) can be equivalently and concisely converted into
\begin{equation}\label{pr:phi_1}
\begin{aligned}
\min_{\boldsymbol{\phi}} ~~ &\frac{\bolds{\xi}_1^H\mathbf{v}\mathbf{v}^H \mathbf{\Xi}_1\mathbf{v}}{\boldsymbol{\phi}^H\mathbf{R}_2\boldsymbol{\phi}}
+\frac{\bolds{\xi}_2^H\mathbf{v}\mathbf{v}^H \mathbf{\Xi}_2\mathbf{v}}{\boldsymbol{\phi}^H\mathbf{R}_1\boldsymbol{\phi}} - \mathbf{v}^H \mathbf{F}\mathbf{v}.
\end{aligned}
\end{equation}
Now, the equivalence between objective functions (\ref{objective_phi_1a}) and (\ref{pr:phi_10}a) is proved.
\end{appendices}

\end{document}